\pdfoutput=1
\documentclass{LMCS}       

\usepackage{amssymb}
\usepackage{amsthm}
\usepackage{latexsym}
\usepackage{cite}
\usepackage{alltt}
\usepackage{enumerate}
\usepackage[pdftitle={CFA2: a Context-Free Approach to Control-Flow Analysis}]{hyperref}
\usepackage{tikz}
\usetikzlibrary{shapes,arrows}

\usepackage{dimvar} 
\usepackage{absint}
\usepackage{cfa2} 
\usepackage{gastex}

\theoremstyle{definition} 
\newtheorem{property}[thm]{Property}

\def\doi{7 (2:3) 2011}
\lmcsheading%
{\doi}
{1--39}
{}
{}
{Jun.~11, 2010}
{May.~\phantom01, 2011}
{}

\begin{document}

\title{CFA2: a Context-Free Approach \\ to Control-Flow Analysis}

\author[D.~Vardoulakis]{Dimitrios Vardoulakis}
\address{Northeastern University} 
\email{\{dimvar,shivers\}@ccs.neu.edu}

\author[O.~Shivers]{Olin Shivers}
\address{\vskip-6 pt} 


\newcommand{\myp}[1]{\tb{#1}}

\begin{abstract}
In a functional language, the dominant control-flow mechanism is function call
and return.
Most higher-order flow analyses, including $k$-CFA, do not handle call and 
return well: they remember only a bounded number of pending calls because they
approximate programs with control-flow graphs.
Call/return mismatch introduces precision-degrading spurious control-flow
paths and increases the analysis time.

We describe CFA2, the first flow analysis with precise call/return matching
in the presence of higher-order functions and tail calls.
We formulate CFA2 as an abstract interpretation of programs in 
continuation-passing style and describe a sound and complete summarization
algorithm for our abstract semantics.
A preliminary evaluation shows that CFA2 gives more accurate data-flow
information than 0CFA and 1CFA.
\end{abstract}

\maketitle

\section*{Introduction}

\noindent Higher-order functional programs can be analyzed using analyses such as the
\kcfa{} family \cite{diss/cmu/91/olin}.
These algorithms approximate the valid control-flow paths through the
program as the set of all paths through a finite graph of abstract
machine states, where each state represents a program point plus some
amount of abstracted environment and control context.

In fact, this is not a particularly tight approximation.
The set of paths through a finite graph is a regular language.
However, the execution traces produced by recursive 
function calls are strings in a \emph{context-free language}.
Approximating this control flow with regular-language techniques
permits execution paths that do not properly match calls with returns.
This is particularly harmful when analyzing higher-order languages, 
since flowing functional values down these spurious paths can 
give rise to further ``phantom'' control-flow structure, 
along which functional values can then flow, and so forth, 
in a destructive spiral that not only degrades precision but
drives up the cost of the analysis.

\emph{Pushdown models} of programs can match an unbounded number of calls and 
returns, tightening up the set of possible executions to strings in a 
context-free language.
Such models have long been used for first-order languages.
The functional approach of Sharir and 
Pnueli~\cite{book/flowanalysis/81/sharir/interproc} computes transfer-functions 
for whole procedures by composing transfer-functions of their basic blocks.
Then, at a call-node these functions are used to compute the data-flow value
of the corresponding return-node directly.
This ``summary-based'' technique has seen widespread 
use~\cite{conf/popl/95/reps/interproc, conf/popl/08/chaudhuri/subcubic}.
Other pushdown models include Recursive State Machines 
\cite{journal/toplas/05/alur/rsm} and Pushdown Systems 
\cite{journal/entcs/97/finkel/pds, conf/concur/97/bouajjani/pds}.

In this paper, we propose \cfat{}, a pushdown model of higher-order 
programs.\footnote{\cfat{} stands for 
``a Context-Free Approach to Control-Flow Analysis''.
We use ``context-free'' with its usual meaning from language theory, to indicate
that \cfat{} approximates valid executions as strings in a context-free 
language.
Unfortunately, ``context-free'' means something else in program analysis.
To avoid confusion, we use ``monovariant'' and ``polyvariant'' when we refer to
the abstraction of calling context in program analysis.
\cfat{} is polyvariant (\aka{} context-sensitive), because it analyzes different
calls to the same function in different environments.
}
Our contributions can be summarized as follows:
\begin{enumerate}[$\bullet$]
\item
  \cfat{} is a flow analysis with precise call/return matching that can be used
  in the compilation of both typed and untyped languages.
  No existing analysis for functional languages enjoys all of these properties.
  \kcfa{} and its variants support lim\-it\-ed call/return matching, bounded by 
  the size of $k$ (section \ref{sec:call-ret-mismatch}).
  Type-based flow analysis with polymorphic subtyping \cite{diss/diku/96/mossin,
    conf/popl/01/rehof/typeflow} also supports limited call/return matching, and
  applies to typed languages only (section \ref{sec:related}).
\item  
  \cfat{} uses a stack and a heap for variable binding.
  Variable references are looked up in one or the other, depending on where they
  appear in the source code.
  Most ref\-er\-ences in typical programs are read from the stack, which results
  in significant pre\-ci\-sion gains. 
  Also, \cfat{} can filter certain bindings off the stack to sharpen 
  pre\-ci\-sion (section~\ref{sec:cfa2sems}).
  \kcfa{} with abstract garbage collection \cite{conf/icfp/06/might/gcfa}
  cannot infer that it is safe to remove these bindings.
  Last, the stack makes \cfat{} resilient to syntax changes like 
  $\eta$-expansion (section \ref{sec:abssems}).
  It is well known that \kcfa{} is sensitive to such changes
  \cite{journal/toplas/98/wright/polysplit, conf/icfp/08/vanhorn/kcfa}.
\item
  We formulate \cfat{} as an abstract interpretation of programs in 
  continuation-passing style (\cps{}).
  The abstract semantics uses a stack of unbounded height.
  Hence, the abstract state space is infinite, unlike \kcfa{}.
  To analyze the state space, we extend the functional approach of Sharir and
  Pnueli \cite{book/flowanalysis/81/sharir/interproc}.
  The resulting algorithm is a search-based variant of summarization that can
  handle higher-order functions and tail recursion.
  Currently, \cfat{} does not handle first-class-control operators such 
  as \tw{call/cc} (section~\ref{sec:summarization}).
\item
  We have implemented \cfa{0}, \cfa{1} and \cfat{} in the Twobit Scheme 
  compiler \cite{conf/lfp/94/clinger/larceny}.
  Our experimental results show that \cfat{} is more precise than \cfa{0} and 
  \cfa{1}.
  Also, \cfat{} usually visits a smaller state space 
  (section~\ref{sec:evaluation}).
\end{enumerate}

\section{Preliminary definitions and notational conventions\label{sec:basics}}

\noindent In flow analysis of \lam-calculus-based languages, a program is usually turned
to an intermediate form where all subexpressions are named before it
is analyzed.  This form can be CPS, administrative normal form
\cite{conf/pldi/93/flanagan/anf}, or ordinary direct-style \lam-calculus where 
each expression has a unique label.
Selecting among these is mostly a matter of taste, and an analysis using one
form can be changed to use another form without much effort.

This work uses CPS.
We opted for CPS because it makes contexts explicit, as 
con\-tin\-u\-a\-tion-lambda terms.
Moreover, \tw{call/cc}, which we wish to support in the future, is directly
expressible in CPS without the need for a special primitive operator.

In this section we describe our \cps{} language.
For brevity, we develop the theory of \cfat{} in the untyped \lam-calculus.
Primitive data, explicit recursion and side-effects can be added using standard
techniques \cite[ch.\ 3]{diss/cmu/91/olin} \cite[ch.\ 9]{diss/07/might/dcfa}.
Compilers that use \cps{} \cite{masters/mit/78/steel/rabbit,%
diss/yale/88/kranz}
usually partition the terms in a program in two disjoint sets, 
the user and the continuation set, 
and treat user terms differently from continuation terms.

We adopt this partitioning for our language (Fig.~\ref{fig:pcps}).
Variables, lambdas and calls get labels from \dulab{} or \dclab.
Labels are pairwise distinct.
User lambdas take a user argument and the current continuation;
continuation lambdas take only a user argument.
We apply an additional syntactic constraint: the only continuation variable
that can appear free in the body of a user lambda \pulam{} is $k$.
This simple constraint forbids first-class 
control~\cite{conf/lfp/92/sabry/cps}.
Intuitively, we get such a program by \cps-converting a direct-style program
without \tw{call/cc}.

\begin{figure}[!t]
  \makebox[\columnwidth][c]{
    \begin{tabular}{c @{\qquad\qquad} c}
      \begin{tabular}{r @{\quad} c @{\quad} l}
        $v \in \dvar$ & $=$ & $\duvar + \dcvar$ \\
        $u \in \duvar$ & $=$ & a set of identifiers \\
        $k \in \dcvar$ & $=$ & a set of identifiers \\

        $\psi \in \dlab$ & $=$ & $\dulab + \dclab$ \\
        $l \in \dulab$ & $=$ & a set of labels \\
        $\gamma \in \dclab$ & $=$ & a set of labels \\

        $\mlam \in \dlam$ & $=$  & $\dulam + \dclam$ \\
        $\mulam \in \dulam$ & \bnf & \denot{\pulam}
      \end{tabular}
      &
      \begin{tabular}{r @{\quad} c @{\quad} l}
        $\mclam \in \dclam$ & \bnf & \denot{\pclam} \\

        $\mcall \in \dcall$ & $=$ & $\ducall +  \dccall$\\
        $\ducall$ & \bnf & \denot{\ucall} \\
        $\dccall$ & \bnf & \denot{\qcall} \\

        $g \in \dexp$ & $=$ & $\duexp + \dcexp$ \\
        $f, e \in \duexp$ & $=$ & $\dulam + \duvar$ \\
        $q \in \dcexp$ & $=$ & $\dclam + \dcvar$ \\

        $\mprog \in \ti{Program}$ & \bnf & \dulam
      \end{tabular}  
    \end{tabular}
  }
  \caption{Partitioned \cps \label{fig:pcps}}
\end{figure}

We assume that all variables in a program have distinct names.
Concrete syntax enclosed in \denot{\cdot} denotes an item of abstract syntax.
Functions with a `?' subscript are predicates, \eg, \isvar{e} returns
true if $e$ is a variable and false otherwise.

We use two notations for tuples, $(e_1, \dots, e_n)$ and 
\tuple{e_1, \dots, e_n}, to avoid confusion when tuples are deeply nested.
We use the latter for lists as well; ambiguities will be resolved by the 
context.
Lists are also described by a head-tail notation, \eg, $3::\tuple{1, 3, -47}$.

\cfat{} treats references to the same variable differently in different 
contexts.
We split references in two categories: stack and heap references.
In direct-style, if a reference appears at the same nesting level as its binder,
then it is a stack reference, otherwise it is a heap reference.
For example, the program \tw{(\ilam{1}(x)(\ilam{2}(y)(x (x y))))} has a stack 
reference to \tw{y} and two heap references to \tw{x}.
Intuitively, only heap references may escape.
When a program $p$ is CPS-converted to a program $p'$, stack (\resp\ heap) 
references in $p$ remain stack (\resp\ heap) references in $p'$.
All references added by the transform are stack references.

We can give an equivalent definition of stack and heap references directly in
CPS, without referring to the original direct-style program.
Labels can be split into disjoint sets according to the innermost user lambda 
that contains them.
In the program \tw{(\ilam{1}(x k1) (k1 (\ilam{2}(y k2) 
(x y (\ilam{3}(u) (x u k2)\(\sp{4}\)))\(\sp{5}\)))\(\sp{6}\))}, 
which is the CPS translation of the previous program, these sets are \mset{1, 6}
and \mset{2, 3, 4, 5}.
The ``label to variable'' map \ltov{\psi} returns all the variables bound by 
any lambdas that belong in the same set as $\psi$, \eg, $\ltov{4} = 
\mset{\tw{y}, \tw{k2}, \tw{u}}$ and $\ltov{6} = \mset{\tw{x}, \tw{k1}}$.
We use this map to model stack behavior, because all continuation lambdas that 
``belong'' to a given user lambda \ilam{l} get closed by extending 
\ilam{l}'s stack frame (\confer{} section~\ref{sec:cfa2sems}).
Notice that, for any $\psi$, \ltov{\psi} contains exactly one continuation 
variable.
Using \ltovNA, we give the following definition.
\vfill\eject

\begin{defi}[Stack and heap references] 
  ~ 
  \begin{enumerate}[$\bullet$]
  \item
    Let $\psi$ be a call site that refers to a variable $v$.
    The predicate \instack{\psi, v} holds iff $v \in \ltov{\psi}$.
    We call $v$ a \tb{stack reference}.
  \item
    Let $\psi$ be a call site that refers to a variable $v$.
    The predicate \inheap{\psi, v} holds iff $v \notin \ltov{\psi}$.
    We call $v$ a \tb{heap reference}.
  \item
    $v$ is a \tb{stack variable}, written \instack{v}, iff all its references 
    satisfy \instackNA.
  \item
    $v$ is a \tb{heap variable}, written \inheap{v}, iff some of its references
    satisfy \inheapNA.
  \end{enumerate}
\end{defi}
\noindent
Then, \instack{5, \tw{y}} holds because $\tw{y} \in 
\mset{\tw{y}, \tw{k2}, \tw{u}}$ and \inheap{5, \tw{x}} holds because
$\tw{x} \notin \mset{\tw{y}, \tw{k2}, \tw{u}}$.

\section{Concrete Semantics}

\begin{figure}[!t]
  {\footnotesize
    \begin{tabular}{@{} l @{$\qquad$} l @{}}
      \lbox{
        \labr{UEA}\; 
        \lbox{
          $(\denot{\ucall}, \cbenv, \cvenv, t) \cstep 
          (\mcproc, \cuarg, \ccarg, \cvenv, l::t)$ \\
          $\mcproc = \cbiga{f, \cbenv, \cvenv}$ \\
          $\cuarg = \cbiga{e, \cbenv, \cvenv}$ \\
          $\ccarg = \cbiga{q, \cbenv, \cvenv}$ 
        }
        \\ \\
        \labr{UAE}\; 
        \lbox{
          $(\mcproc, \cuarg, \ccarg, \cvenv, t) \cstep 
          (\mcall, \cbenv', \cvenv', t)$ \\
          $\mcproc \equiv \tuple{\denot{\ulam}, \cbenv}$ \\
          $\cbenv' = \cbenv \onemap{u}{t} \onemap{k}{t}$ \\
          $\cvenv' = \cvenv \onemap{(u, t)}{\cuarg} \onemap{(k, t)}{\ccarg}$
        }
        \\ \\
        \labr{CEA}\; 
        \lbox{
          $(\denot{\qcall}, \cbenv, \cvenv, t) \cstep 
          (\mcproc, \cuarg, \cvenv, \gamma::t)$ \\
          $\mcproc = \cbiga{q, \cbenv, \cvenv}$ \\
          $\cuarg = \cbiga{e, \cbenv, \cvenv}$ 
        }
        \\ \\
        \labr{CAE}\; 
        \lbox{
          $(\mcproc, \cuarg, \cvenv, t) \cstep (\mcall, \cbenv', \cvenv', t)$ \\
          $\mcproc = \tuple{\denot{\clam}, \cbenv}$\\
          $\cbenv' = \cbenv \onemap{u}{t}$ \\
          $\cvenv' = \cvenv \onemap{(u, t)}{\cuarg}$ 
        }
      }
      &
      \lbox{
        \raisebox{-2ex}{
          $\quad \cbiga{g, \cbenv, \cvenv} \,\triangleq\,
          \begin{cases}
            (g, \cbenv) & \islam{g} \\
            \cvenv(g, \cbenv(g)) & \isvar{g}
          \end{cases}$
        }
        \\ \\
        \begin{tabular}{@{} r @{\;} c @{\;} l @{}}
          \multicolumn{3}{c}{\hspace*{\stretch{1}}
            Concrete domains:\hspace*{\stretch{6}}} \rule{0cm}{0.4cm} \\
          $\cstat \in \dstate$ & $=$  & $\deval + \dapply$ 
          \rule{0cm}{0.4cm} \\ 
          $\deval$ & $=$  & $\dueval + \dceval$ \\
          $\dueval$ & $=$  & $\ducall \times \dbenv \times \dvenv
          \times \dtime$ \\
          $\dceval$ & $=$  & $\dccall \times \dbenv \times \dvenv
          \times \dtime$  \\
          $\dapply$ & $=$  & $\duapply + \dcapply$ \\
          $\duapply$ & $=$  & 
          $\duclos \times \!\duclos \times \!\dcclos \times \!\dvenv 
          \times \!\dtime$ \\
          $\dcapply$ & $=$ & 
          $\dcclos \times  \duclos \times \!\dvenv \times \!\dtime$ \\
          $\dclos$ & $=$ & $\duclos + \dcclos$ \\
          $\cuarg \in \duclos$ & $=$ & $\dulam \times \dbenv$ \\
          $\ccarg \in \dcclos$ & $=$  & $(\dclam \times \dbenv) + \haltcont$ \\
          $\cbenv \in \dbenv$ & $=$  & $\dvar \rightharpoonup \dtime$ \\
          $\cvenv \in \dvenv$ & $=$  & 
          $\dvar \times \dtime \rightharpoonup \dclos$\\
          $t \in \dtime$ & $=$  & $\dlab^*$ 
        \end{tabular}
      }
    \end{tabular}
  }
\caption{Concrete semantics and domains for Partitioned CPS\label{fig:concsems}}
\end{figure}

\noindent Execution in Partitioned CPS is guided by the semantics of 
Fig.~\ref{fig:concsems}.
In the terminology of abstract interpretation, this semantics is called the
\emph{concrete} semantics.
In order to find properties of a program at compile time, one needs to derive a
computable approximation of the concrete semantics, called the \emph{abstract}
semantics.
\cfat{} and \kcfa{} are such approximations.

Execution traces alternate between \deval{} and \dapply{} states.
At an \deval{} state, we evaluate the subexpressions of a call site before
performing a call.
At an \dapply, we perform the call.

The last component of each state is a \emph{time}, which is a sequence of 
call sites.
\deval{} to \dapply{} transitions increment the time by recording the label
of the corresponding call site.
\dapply{} to \deval{} transitions leave the time unchanged.
Thus, the time $t$ of a state reveals the call sites along the execution path
to that state.

Times indicate points in the execution when variables are bound.
The binding environment \cbenv{} is a partial function that maps variables 
to their binding times.
The variable environment \cvenv{} maps variable-time pairs to values.
To find the value of a variable $v$, we look up the time $v$ was put in \cbenv, 
and use that to search for the actual value in \cvenv.

Let's look at the transitions more closely.
At a \dueval{} state with call site \ucall, we evaluate $f$, $e$ and $q$
using the function \cbigaNA.
Lambdas are paired up with \cbenv{} to become closures, while variables are 
looked up in \cvenv{} using \cbenv.
We add the label $l$ in front of the current time and transition to a 
\duapply{} state (rule \labr{UEA}).

From \duapply{} to \deval{}, we bind the formals of a procedure 
\tuple{\denot{\ulam}, \cbenv} to the arguments and jump to its body.
The new binding environment $\cbenv'$ extends the procedure's environment, with
$u$ and $k$ mapped to the current time.
The new variable environment $\cvenv'$ maps $(u, t)$ to the user argument 
\cuarg, and $(k, t)$ to the continuation \ccarg{} (rule \labr{UAE}).

The remaining two transitions are similar.
We use \haltcont{} to denote the top-level continuation of a program \mprog.
The initial state \initstate{} is 
$((\mprog,\emptyset), \ti{input}, \haltcont, \emptyset, \tuple{})$, 
where \ti{input} is a closure of the form $\tuple{\denot{\ulam},\emptyset}$.
The initial time is the empty sequence of call sites.

\cps{}-based compilers may or may not use a stack for the final code.
Steele's view, illustrated in the Rabbit compiler 
\cite{masters/mit/78/steel/rabbit}, is that argument evaluation pushes stack and
function calls are {\small GOTO}s.
Since arguments in \cps{} are not calls,
argument evaluation is trivial and Rabbit never needs to push stack.
By this approach, every call in \cps{} is a tail call.

An alternative style was used in the Orbit compiler \cite{diss/yale/88/kranz}.
At every function call, Orbit pushes a frame for the arguments.
By this approach, tail calls are only the calls where the continuation 
argument is a variable.
These \cps{} call sites were in tail position in the initial direct-style 
program.
\dceval{} states where the operator is a variable are calls to the current 
continuation with a return value.
Orbit pops the stack at tail calls and before calling the current continuation.

We will see later that the abstract semantics of \cfat{} uses a stack, 
like Orbit.
However, \cfat{} computes safe flow information which can be used by both 
aforementioned approaches.
The workings of the abstract interpretation are independent of what style
an implementor chooses for the final code.

\section{Limitations of \kcfa\label{sec:kcfa}}

\noindent In this section, we discuss the main causes of imprecision and inefficiency 
in \kcfa.
Our motivation in developing \cfat{} is to create an analysis that overcomes 
these limitations.

We assume some familiarity with \kcfa, and abstract interpretation in general.
Detailed descriptions on these topics can be found in
\cite{diss/cmu/91/olin, diss/07/might/dcfa}.
We use Scheme syntax for our example programs.

\subsection{{\kcfa} does not properly match calls and 
returns\label{sec:call-ret-mismatch}}

\begin{figure}[!t]
  \begin{tabular}{@{} l @{} r @{}}
    \begin{tabular}[b]{@{} l @{}}
      \begin{minipage}[b]{0.35\linewidth}
        \begin{alltt}
(define (len l)
  (if (pair? l)
      (+ 1 (len (cdr l)))
      0))
(len '(3)) \end{alltt}
$\qquad\qquad\Downarrow$ {\footnotesize CPS}
\begin{alltt}
(define (len l k)
 (pair? l)
  (\lam(test)
   (if test
    (\lam() 
     (cdr l
      (\lam(rest)
        (len rest
         (\lam(ans) 
           (+ 1 ans k))))))
    (\lam() (k 0)))))
(len '(3) \haltcont) \end{alltt}
      \end{minipage}
  \end{tabular}
  &
  \begin{minipage}[b]{0.65\linewidth}
  \begin{tikzpicture}[scale=0.5, node distance=1.3cm, ->, >=stealth']
    \tikzset{
      font={\ttfamily \footnotesize},
      entry/.style = { rectangle, draw },
      exit/.style = { rectangle, double, draw },
      inner/.style = { rectangle, rounded corners, draw },
      callret/.style = {
        rectangle split,
        rectangle split parts = 2,
        rounded corners,
        draw
      }
    }

    \node (auxMain) {};
    \node (inMain) [entry, below of=auxMain, node distance=2.5cm, label=left:1]
    {main()};
    \node (lencall) [callret, below of=inMain, label=173:2, label=187:3]
    {len '(3) \nodepart{second} ret};
    \node (outMain) [exit, below of=lencall, label=left:4]
    {main};

    \draw [->] (inMain) -- (lencall);
    \draw [->] (lencall) -- (outMain);

    \node (inLen) [entry, label=176:5, right of = auxMain, node distance=5cm] 
    {len(l)};
    \node (len2) [inner, below of = inLen, label=left:6] {test := pair?~l};
    \node (isPair) [inner, below of = len2, label=left:7] {test};
    \node (auxlen) [below of = isPair] {};
    \node (tru) [inner, right of = auxlen, label=left:9] {rest := cdr l};
    \node (recur) [callret, below of = tru, label=173:10, label=187:11]
    {len rest \nodepart{second} ans};
    \node (afterrecur) [inner, below of = recur, label=left:12]
    {ret := 1 + ans};
    \node (fals) [inner, left of = auxlen, left of = recur, label=left:8]
    {ret := 0};
    \node (auxlen2) [left of = afterrecur] {};
    \node (outLen) [exit, below of = auxlen2, label=160:13] {len};

    \draw [->] (inLen) -- (len2);
    \draw [->] (len2) -- (isPair);
    \path (isPair) edge node{\#f$\phantom{abcd}$} (fals);
    \path (isPair) edge node{$\phantom{abcd}$\#t} (tru);
    \draw [->] (fals) -- (outLen);
    \draw [->] (tru) -- (recur);
    \draw [->] (recur) -- (afterrecur);
    \draw [->] (afterrecur) -- (outLen);

    \draw [->, dashed] (recur.north east) to[in=0, right=2cm] (inLen.east);
    \draw [->, dashed] 
    (outLen.east) to[out=0, in=-5, right=5cm] (recur.south east);
    \draw [->, dashed] (lencall.north east) to[bend left] (inLen);
    \draw [->, dashed] (outLen.west) to[bend left] (lencall.south east);
  \end{tikzpicture}
  \end{minipage}
  \end{tabular}
\caption{\cfa{0} on \tw{len} \label{fig:0cfa-eg}}
\end{figure}

In order to make the state space of \kcfa{} finite, Shivers chose a mechanism 
similar to the call-strings of Sharir and 
Pnueli \cite{book/flowanalysis/81/sharir/interproc}.
Thus, recursive programs introduce approximation by folding an unbounded number
of recursive calls down to a fixed-size call-string.
In effect, by applying \kcfa{} to a higher-order program, we turn it into a 
finite-state machine.
Taken to the extreme, when $k$ is zero, a function can return to any of its
callers, not just to the last one. 

For example, consider the function \tw{len} that computes the length of a list.
Fig.~\ref{fig:0cfa-eg} shows the code for \tw{len}, its CPS translation and the
associated control-flow graph.
In the graph, the top level of the program is presented as a function called 
\tw{main}.
Function entry and exit nodes are rectangles with sharp corners.
Inner nodes are rectangles with round\-ed corners.
Each call site is represented by a call node and a corresponding return node,
which contains the variable to which the result of the call is assigned.
Each function uses a local variable \tw{ret} for its return value.
Solid arrows are intraprocedural steps.
Dashed arrows go from call sites to function entries and from function exits to
return points.
There is no edge between call and return nodes; a call reaches its corresponding
return only if the callee terminates.
A monovariant analysis, such as \cfa{0}, considers every path from 1 to 4 to be
a valid execution.
In particular, it cannot exclude the path 1, 2, 5, 6, 7, 9, 10, 5, 6, 7, 8, 13,
3, 4.
By following such a path, the program will terminate with a \emph{non-empty}
stack.
It is clear that {\kcfa} cannot help much with optimizations that require
accurate cal\-cu\-la\-tion of the stack change between program states, such as 
stack allocation of closure environments.

Spurious flows caused by call/return mismatch affect traditional data-flow
information as well.
For instance, \cfa{0}-constant-propagation for the program below
cannot spot that \tw{n2} is the constant 2,
because 1 also flows to \tw{x} and is mistakenly returned by the second call to
\tw{app}.
\cfa{1} also fails, because both calls to \tw{id} happen in the body of 
\tw{app}.
\cfa{2} helps in this example, but repeated $\eta$-expansion of \tw{id} can 
trick \kcfa{} for any $k$.
\begin{center}
  \begin{minipage}{0.45\columnwidth}
    \begin{alltt}
(let* ((app (\lam(f e) (f e)))
       (id (\lam(x) x))
       (n1 (app id 1))
       (n2 (app id 2)))
  (+ n1 n2)) \end{alltt}    
  \end{minipage}
\end{center}

In a non-recursive program, a large enough $k$ can provide accurate 
call/return matching, but this is not desirable because the analysis becomes
intractably slow even when $k$ is 1~\cite{conf/icfp/08/vanhorn/kcfa}.
Moreover, the ubiquity of recursion in functional programs calls for a static
analysis that can match an unbounded number of calls and returns.
This can be done if we approximate programs using pushdown models instead of
finite-state machines.

\subsection{The environment problem and fake rebinding\label{sec:fake-rebind}}

In higher-order languages, many bindings of the same variable can be 
simultaneously live.
Determining at compile time whether two references to some variable will be 
bound in the same run-time environment is referred to as 
the \emph{environment problem}~\cite{diss/cmu/91/olin}.
Consider the following program:
\begin{center}
  \begin{minipage}{0.8\columnwidth}
    \begin{alltt}
(let ((f (\lam(x thunk) (if (number? x) (thunk) (\ilam{1}() x)))))
  (f 0 (f "foo" "bar"))) \end{alltt}    
  \end{minipage}
\end{center}
In the inner call to \tw{f}, \tw{x} is bound to \tw{"foo"} and \ilam{1} is 
returned.
We call \tw{f} again; this time, \tw{x} is 0, so we jump through \tw{(thunk)} to
\ilam{1}, and reference \tw{x}, which, despite the just-completed 
test, is \emph{not} a number: it is the string \tw{"foo"}.
Thus, during abstract interpretation, it is gen\-er\-al\-ly \emph{unsafe} to 
assume that a reference has some property just because an ear\-li\-er reference
had that property.
This has an unfortunate consequence: sometimes an ear\-li\-er reference provides
\emph{safe} information about the reference at hand and \kcfa{} does not spot 
it:
\begin{center}
  \begin{minipage}{0.55\columnwidth}
    \begin{alltt}
(define (compose-same f x) (f (f x)\icsite{1})\icsite{2}) \end{alltt}    
  \end{minipage}
\end{center}
In \tw{compose-same}, both references to \tw{f} are always bound at the same 
time.
However, if multiple closures flow to \tw{f}, {\kcfa} may call one closure at
call site 1 and a different closure at call site 2.
This flow never happens at run time.

Imprecise binding information also makes it difficult to infer the types of
variable references.
In \tw{len}, the \tw{cdr} primitive must perform a run-time check and signal an
error if \tw{l} is not bound to a pair.
This check is redundant since we checked for \tw{pair?}\ earlier, and both 
ref\-er\-ences to \tw{l} are bound in the same environment.
If \tw{len} is called with both pair and non-pair arguments, \kcfa{} cannot
eliminate the run-time check.

\cfat{} tackles this problem by distinguishing stack from heap references.
If a reference $v$ appears in a static context where we know the current stack 
frame is its environment record, we can be precise.
If $v$ appears free in some possibly escaping lambda, we cannot predict its
extent so we fall back to a conservative approximation.

\subsection{Imprecision increases the running time of the analysis}

\kcfa{} for $k > 0$ is not a cheap analysis, both in theory
\cite{conf/icfp/08/vanhorn/kcfa} and in practice
\cite{conf/pldi/88/shivers/cfaretro}.
Counterintuitively, imprecision in higher-order flow analyses can increase 
their running time: imprecision induces spurious control paths, along which the
analysis must flow data, thus creating further spurious paths, and so on, in a 
vicious cycle which creates extra work whose only function is to degrade 
pre\-ci\-sion.
This is why techniques that aggressively prune the search space, such as 
\cfa{\Gamma}~\cite{conf/icfp/06/might/gcfa}, not only increase precision, but 
can also improve the speed of the analysis.

In the previous subsections, we saw examples of information known at compile 
time that \kcfa{} cannot exploit.
\cfat{} uses this information. 
The enhanced precision of \cfat{} has a positive effect on its running
time (\confer{} section~\ref{sec:evaluation}).

\section{The {\cfat} semantics\label{sec:cfa2sems}}

\begin{figure}[!t]
  {\footnotesize
    \begin{tabular}{@{} l r @{}}
      \lbox{
        \labar{UEA}
        \lbox{
          $(\denot{\ucall}, \stenv, \henv) \astep 
          (\mulam, \auarg, \acarg, \stenv', \henv)$ \\
          $\mulam \in \ubiga{f, l, \stenv, \henv}$ \\
          $\auarg = \ubiga{e, l, \stenv, \henv}$ \\
          $\acarg = \kbiga{q, \stenv}$ \\
          $\stenv' \!=\! 
          \begin{cases}
            \pop{\stenv} & \isvar{q} \\
            \stenv & 
            \!\!\!\!\!\!\!\!\!\!\!\!\!\!\!\!\!\!\!\!\!\!\!\!\!\!\!\!\!
            \islam{q} \land (\inheap{l, f} \lor \islam{f}) \\
            \stenv\onemap{f\!}{\!\mset{\mulam}} & \islam{q} \land \instack{l, f}
          \end{cases}$ 
        } \\ \\
        \labar{UAE}
        \lbox{
          $(\denot{\ulam}, \auarg, \acarg, \stenv, \henv) \astep 
          (\mcall, \stenv', \henv')$ \\
          $\stenv' = \push{\onemap{u}{\auarg} \onemap{k}{\acarg}} {\stenv}$ \\
          $\henv' = 
          \begin{cases}
            \henv\join\onemap{u}{\auarg} & \inheap{u} \\
            \henv & \instack{u}
          \end{cases}$ 
        } \\ \\
        \labar{CEA}
        \lbox{
          $(\denot{\qcall}, \stenv, \henv) \astep 
          (\mclam, \auarg, \stenv', \henv)$ \\
          $\mclam = \kbiga{q, \stenv}$ \\
          $\auarg = \ubiga{e, \gamma, \stenv, \henv}$ \\
          $\stenv' = 
          \begin{cases}
            \pop{\stenv} & \isvar{q} \\
            \stenv & \islam{q}
          \end{cases}$
        } \\ \\
        \labar{CAE}
        \lbox{
          $(\denot{\clam}, \auarg, \stenv, \henv) \astep 
          (\mcall, \stenv', \henv')$ \\
          $\stenv' = \stenv\onemap{u}{\auarg}$ \\
          $\henv' = 
          \begin{cases}
            \henv\join\onemap{u}{\auarg} & \inheap{u} \\
            \henv & \instack{u}
          \end{cases}$
        }
      }
      & 
      \lbox{
        \raisebox{-1cm}{
          \begin{tabular}{@{} p{0.1cm} r @{}}
            ~ &
            \begin{tabular}{@{} r @{\,} c @{\,} l @{}}
              \ubiga{e, \psi, \stenv, \henv} 
              & $\triangleq$ &
              $\begin{cases}
                \mset{e} & \islam{e} \\
                \stenv(e) & \instack{\psi, e} \\
                \henv(e) & \inheap{\psi, e}
              \end{cases}$
              \\ \\
              \kbiga{q, \stenv}
              & $\triangleq$ &
              $\begin{cases}
                q & \islam{q} \\
                \stenv(q) & \isvar{q} 
              \end{cases}$
            \end{tabular}
          \end{tabular}
        }
        \\ \\
        \begin{tabular}{@{} r @{\,} c @{\:} l @{}}
          \multicolumn{3}{c}{Abstract domains:\makebox[2.5cm][t]{}}
          \\
          $\astat \in \daueval$ & $=$ & $\ducall \times \dstack \times \dheap$ 
          \rule{0cm}{0.45cm} \\
          $\astat \in \dauapply$ & $=$ & 
          $\dulam \!\times\! \dauclos \!\times\! \dacclos \!\times\!
          \dstack \!\times\! \dheap$
          \\
          $\astat \in \daceval$ & $=$ & $\dccall \times \dstack \times \dheap$ 
          \\
          $\astat \in \dacapply$ & $=$ & 
          $\dacclos \times \dauclos \times \dstack \times \dheap$
          \\
          $\auarg \in \dauclos$ & $=$ &  $\powerset{\dulam}$ 
          \\
          $\acarg \in \dacclos$ & $=$ & $\dclam + \haltcont$
          \\
          $\frenv, \tfenv \in \dframe$ & $=$ &
          $(\duvar \!\rightharpoonup\! \dauclos) \,\cup
          (\dcvar \!\rightharpoonup\! \dacclos)$ 
          \\
          $\stenv \in \dstack$ & $=$ & $\dframe^*$ 
          \\
          $\henv \in \dheap$ & $=$ & $\duvar \rightharpoonup \dauclos$
        \end{tabular}
        \\ \\
        \begin{tabular}{@{$\quad$} r @{\;\;} p{0.1cm} @{\;\;\;} l}
          \multicolumn{3}{c}{Stack operations:}
          \\ 
          $\pop{\tfenv::\stenv}$ & $\triangleq$ & $\stenv$ \rule{0cm}{0.45cm}
          \\ 
          $\push{\frenv}{\stenv}$ & $\triangleq$ & $\frenv::\stenv$ 
          \rule{0cm}{0.45cm}
          \\ 
          $(\tfenv::\stenv)(v)$ & $\triangleq$ & $\tfenv(v)$ \rule{0cm}{0.45cm}
          \\ 
          $(\tfenv::\stenv)\onemap{u}{\auarg}$ & $\triangleq$ &
          $\tfenv\onemap{u}{\auarg}\!::\!\stenv$ \rule{0cm}{0.45cm}      
        \end{tabular}
      }
    \end{tabular}
  }  
  \caption{Abstract semantics and relevant definitions \label{fig:abssems}}
\end{figure}

\noindent In this section we define the abstract semantics of \cfat.
The abstract semantics approximates the concrete semantics.
This means that each concrete state has a corresponding abstract state.
Therefore, each concrete execution, \ie, sequence of states related by \cstep,
has a corresponding abstract execution that computes an approximate answer.

Each abstract state has a stack.
Analyzing recursive programs requires states with stacks of unbounded size.
Thus, the abstract state space is infinite and the standard algorithms for 
\kcfa{} \cite{diss/cmu/91/olin, diss/07/might/dcfa} will diverge because they
work by enumerating all states.
We show how to solve the stack-size problem in section~\ref{sec:summarization}.
Here, we describe the abstract semantics (section \ref{sec:abssems}), show how
to map concrete to abstract states and prove the correctness of the abstract
semantics (section \ref{sec:simulation}).

\subsection{Abstract semantics\label{sec:abssems}}

The \cfat{} semantics is an abstract interpreter that executes a CPS program,
using a stack for variable binding and return-point information.

We describe the stack-management policy with an example.
Assume that we run the \tw{len} program of section~\ref{sec:kcfa}.
When calling \tw{(len '(3) \haltcont)} we push a frame 
$[\tw{l} \mapsto \tw{(3)}][\tw{k} \mapsto \haltcont]$ on the stack.
The test \tw{(pair?} \tw{l)} is true, so we add the binding 
$[\tw{test} \mapsto \ti{true}]$ to the top frame and jump to the true branch.
We take the \tw{cdr} of \tw{l} and add the binding $[\tw{rest} \mapsto \tw{()}]$
to the top frame.
We call \tw{len} again, push a new frame for its arguments and jump to its body.
This time the test is false, so we extend the top frame with
$[\tw{test} \mapsto \ti{false}]$ and jump to the false branch.
The call to \tw{k} is a function return, so we pop a frame and pass \tw{0} 
to \tw{(\lam(ans)(+ 1 ans k))}.
Call site \tw{(+ 1 ans k)} is also a function return, so we pop the remaining 
frame and pass \tw{1} to the top-level continuation \haltcont.

In general, we push a frame at function entries and pop at tail calls and at
function returns.
Results of intermediate computations are stored in the top frame.
This policy enforces two invariants about the abstract interpreter.
First, when executing inside a user function \pulam, the domain of the top frame
is a subset of \ltov{l}.
Second, the frame below the top frame is the environment of the current
continuation.

Each variable $v$ in our example was looked up in the top frame,
because each lookup happened while executing inside the lambda that binds $v$.
This is not always the case; in the first snippet of section
\ref{sec:fake-rebind} there is a heap reference to \tw{x} in \ilam{1}.
When control reaches that reference, the top frame does not belong to the
lambda that binds \tw{x}.
In \cfat, we look up stack references in the top frame, and heap references in 
the heap.
Stack lookups below the top frame never happen.

The \cfat{} semantics appears in Fig.~\ref{fig:abssems}.
An abstract value is either an abstract user closure (member of the 
set \dauclos) or an abstract continuation closure (member of \dacclos).
An abstract user closure is a set of user lambdas.
An abstract continuation closure is either a continuation lambda or \haltcont.
A frame is a map from variables to abstract values, 
and a stack is a sequence of frames.
All stack operations except \pushNA{} are defined for non-empty stacks only.
A heap is a map from variables to abstract values.
It contains only user bindings because, without first-class control, every 
continuation variable is a stack variable.

On transition from a \daueval{} state \astat{} to a \dauapply{} state $\astat'$
(rule \labar{UEA}), we first evaluate $f$, $e$ and $q$.
We evaluate user terms using \ubigaNA{} and continuation terms using \kbigaNA.
We non-deterministically choose one of the lambdas that flow to $f$ as the
operator in $\astat'$.\footnote{An abstract execution explores one path, but the
algorithm that searches the state space considers all possible executions
(\confer{} section \ref{sec:summarization}), as is the case in the operational
formulation of \kcfa~\cite{diss/07/might/dcfa}.}
The change to the stack depends on $q$ and $f$.
If $q$ is a variable, the call is a tail call so we pop the stack (case 1).
If $q$ is a lambda, it evaluates to a new closure whose environment is the top 
frame, hence we do not pop the stack (cases 2, 3).
Moreover, if $f$ is a lambda or a heap reference then we leave the
stack unchanged.
However, if $f$ is a stack reference, we set $f$'s value in the top 
frame to \mset{\mulam}, possibly forgetting other lambdas that flow to $f$.
This ``stack filtering'' prevents fake 
rebinding ({\confer} section~\ref{sec:fake-rebind}):
when we return to \acarg{}, we may reach more stack references of $f$.
These references and the current one are bound at the same time.
Since we are committing to \mulam{} in this transition, 
these references must also be bound to \mulam.

In the \dauapply-to-\daeval{} transition (rule \labar{UAE}), we push a frame for
the procedure's arguments.
In addition, if $u$ is a heap variable we must update its binding in the heap.
The join operation \join{} is defined as:
\begin{center}
  $(h \join \onemap{u}{\auarg})(v) \triangleq
  \begin{cases}
    h(v) & v \not\equiv u \\
    h(v) \cup \auarg{} & v \equiv u
  \end{cases}$
\end{center}

In a \daceval-to-\dacapply{} transition (rule \labar{CEA}), we are preparing for
a call to a continuation so we must reset the stack to the stack of its birth. 
When $q$ is a variable, the \daceval{} state is a function return and the
continuation's environment is the second stack frame.
Therefore, we pop a frame before calling \mclam.
When $q$ is a lambda, 
it is a newly created closure thus the stack does not change.
Note that the transition is deterministic, unlike \labar{UEA}.
Since we always know which continuation we are about to call, call/return
mismatch \emph{never} happens.
For instance, the function \tw{len} may be called from many places in a program,
so multiple continuations may flow to \tw{k}.
But, by retrieving \tw{k}'s value from the stack, we always return to the 
correct continuation.

In the \dacapply-to-\daeval{} transition (rule \labar{CAE}), our stack policy
dictates that we extend the top frame with the binding for the continuation's
parameter $u$. 
If $u$ is a heap variable, we also update the heap.\footnote{All temporaries
created by the \cps{} transform are stack variables; but a compiler 
optimization may rewrite a program to create heap references to temporaries.}

\paragraph{\myp{Examples}}
When the analyzed program is not recursive, the stack size is bounded so we can
enumerate all abstract states without diverging.
Let's see how the abstract semantics works on a simple program that applies the
identity function twice and returns the result of the second call.
The initial state \ainitstate{} is a \dauapply.
{\newcommand{\callh}{\tw{(h n2)}}
  \newcommand{\lamii}{\tw{(\ilam{2}(n2)\callh)}}
  \newcommand{\callii}{\tw{(id 2 \lamii)}}
  \newcommand{\lami}{\tw{(\ilam{1}(n1)\callii)}}
  \newcommand{\calli}{\tw{(id 1 \lami)}}
  \newcommand{\lamtop}{\tw{(\lam(id h)\calli)}}
  \newcommand{\callid}{\tw{(k x)}}
  \newcommand{\lamid}{\tw{(\ilam{3}(x k)\callid)}}
  \newcommand{\framei}{\onemap{\tw{id}}{\mset{\ilam{3}}}
    \onemap{\tw{h}}{\haltcont}}
  \newcommand{\frameii}{\onemap{\tw{n1}}{\mset{1}}\framei}
  \newcommand{\frameiii}{\onemap{\tw{n2}}{\mset{2}}\frameii}
  \newcommand{\frameidi}{\onemap{\tw{x}}{\mset{1}}\onemap{\tw{k}}{\ilam{1}}}
  \newcommand{\frameidii}{\onemap{\tw{x}}{\mset{2}}\onemap{\tw{k}}{\ilam{2}}}
  $$(\denot{\lamtop}, \mset{\denot{\lamid}}, \haltcont, \tuple{}, \emptyset)$$ 
  All variables in this example are stack variables, so the heap will remain
  empty throughout the execution.
  In frames, we abbreviate lambdas by their labels.
  By rule \labar{UAE}, we push a frame for \tw{id} and \tw{h} and transition to
  a \daueval{} state.
  $$(\denot{\calli}, \tuple{\framei}, \emptyset)$$
  We look up \tw{id} in the top frame.
  Since the continuation argument is a lambda, we do not pop the stack.
  The next state is a \dauapply.
  $$(\denot{\lamid}, \mset{1}, \ilam{1}, \tuple{\framei}, \emptyset)$$
  We push a frame for the arguments of \ilam{3} and jump to its body.
  $$(\denot{\callid}, \tuple{\frameidi, \framei}, \emptyset)$$
  This is a \daceval{} state where the operator is a variable, so we pop a 
  frame.
  $$(\denot{\lami}, \mset{1}, \tuple{\framei}, \emptyset)$$
  We extend the top frame to bind \tw{n1} and jump to the body of \ilam{1}.
  $$(\denot{\callii}, \tuple{\frameii}, \emptyset)$$
  The new call to \tw{id} is also not a tail call, so we do not pop.
  $$(\denot{\lamid}, \mset{2}, \ilam{2}, \tuple{\frameii}, \emptyset)$$
  We push a frame and jump to the body of \ilam{3}.
  $$(\denot{\callid}, \tuple{\frameidii, \frameii}, \emptyset)$$
  We pop a frame and jump to \ilam{2}.
  $$(\denot{\lamii}, \mset{2}, \tuple{\frameii}, \emptyset)$$
  We extend the top frame to bind \tw{n2} and jump to the body of \ilam{2}.
  $$(\denot{\callh}, \tuple{\frameiii}, \emptyset)$$
  The operator is a variable, so we pop the stack.
  The next state is a final state, so the program terminates with value 
  \mset{2}.
  $$(\haltcont, \mset{2}, \tuple{}, \emptyset)$$
}%
\cfa{1} would also find the precise answer for this program.
However, if we $\eta$-expand \ilam{3} to 
\tw{(\ilam{3}(x k)((\ilam{4}(y k2)(k2 y)) x k))}, \cfa{1} will return \mset{1,2}
because both calls to \ilam{4} happen at the same call site.
\cfat{} is more resilient to $\eta$-expansion.
It will return the precise answer in the modified program because the change did
not create any heap references.
However, if we change \ilam{3} to
\tw{(\ilam{3}(x k)((\ilam{4}(y k2)(k2 x)) x k))},
then both 1 and 2 flow to the heap reference to \tw{x} and \cfat{} will return
\mset{1, 2}.

\subsection{Correctness of the abstract semantics\label{sec:simulation}}

\begin{figure}[!t]
  {\footnotesize
    \renewcommand{\arraystretch}{2}
    \lbox{
      $\ctoa{(\denot{\slp g_1 \dots g_n\srp^{\psi}}, \cbenv, \cvenv, t)} = 
      (\denot{\slp g_1 \dots g_n\srp^{\psi}}, \: 
      \tostack{\ltov{\psi}}{\cbenv}{\cvenv}, \: \ctoa{\cvenv} )$ 
      \\ \renewcommand{\arraystretch}{1}\!\!
      \lbox{
        $\ctoa{(\tuple{\denot{\ulam}, \cbenv}, \cuarg, \ccarg, \cvenv, t)} = 
        (\denot{\ulam}, \ctoa{\cuarg}, \ctoa{\ccarg}, \stenv, \ctoa{\cvenv})$ 
        \rule{0cm}{0.47cm} \\
        where 
        $\stenv =
        \begin{cases}
          \tuple{} & \ccarg = \haltcont \\
          \tostack{\ltov{\gamma}}{\cbenv'}{\cvenv} &
          \ccarg = (\denot{\slp\lambda_{\gamma}(u') \mcall '\srp} , \cbenv') 
        \end{cases}$ 
      }
      \\ \renewcommand{\arraystretch}{2}\!\!
      $\ctoa{(\tuple{\denot{\clam}, \cbenv}, \cuarg, \cvenv, t)} = 
      (\denot{\clam}, \,\ctoa{\cuarg}, \tostack{\ltov{\gamma}}{\cbenv}{\cvenv},
      \,\ctoa{\cvenv})$ \rule{0cm}{0.65cm}
      \\
      $\ctoa{(\haltcont, \cuarg, \cvenv, t)} = 
      (\haltcont, \ctoa{\cuarg}, \,\tuple{}, \,\ctoa{\cvenv})$
      \\
      $\ctoa{(\denot{\ulam}, \cbenv)} = \mset{\denot{\ulam}}\rule{0cm}{0.47cm}$ 
      \\
      $\ctoa{(\denot{\clam}, \cbenv)} = \denot{\clam}{}$ 
      \\
      $\ctoa{\haltcont} = \haltcont$ 
      \\
      $\ctoa{\cvenv} = 
      \msetcomp{(u, \bigsqcup_{t}\: \ctoa{\cvenv(u, t)})}{\inheap{u}}$ 
      \\ \renewcommand{\arraystretch}{1}\!\!
      \lbox{
        $\tostack{\mset{u_1, \dots, u_n, k}}{\cbenv}{\cvenv} \triangleq
        \begin{cases}
          \tuple{\manymap{u_i}{\auarg_i} \onemap{k}{\haltcont}} & 
          \eh{\haltcont = \cvenv(k, \cbenv(k))} 
          \\
          \manymap{u_i}{\auarg_i} \onemap{k}{\denot{\pclam}} :: \stenv 
          & 
          \eh{(\denot{\clam}, \cbenv') = \cvenv(k, \cbenv(k))} \rule{0cm}{0.6cm}
        \end{cases}$ \rule{0cm}{0.9cm} \\
        where $\auarg_i = \ctoa{\cvenv(u_i, \cbenv(u_i))}\;$ and 
        $\;\stenv = \tostack{\ltov{\gamma}}{\cbenv'\!}{\cvenv}$\rule{0cm}{0.4cm}
      }
    }
  }
  \caption{From concrete states to abstract states \label{fig:conc-to-abs}}
\end{figure}

\begin{figure}[!t]
  {\footnotesize
    {\renewcommand{\arraystretch}{1.7}
      \lbox{
        $(\mcall, \stenv_1, \henv_1) \sqsubseteq (\mcall, \stenv_2, \henv_2)$
        \quad iff \quad
        $\stenv_1 \sqsubseteq \stenv_2 \;\land\; \henv_1 \sqsubseteq \henv_2$ 
        \\
        $(\mulam, \auarg_1, \acarg, \stenv_1, \henv_1) \sqsubseteq 
        (\mulam, \auarg_2, \acarg, \stenv_2, \henv_2)$
        \quad iff \quad
        $\auarg_1 \sqsubseteq \auarg_2 \;\land\; \stenv_1 \sqsubseteq \stenv_2 
        \;\land\; \henv_1 \sqsubseteq \henv_2$
        \\
        $(\acarg, \auarg_1, \stenv_1, \henv_1) \sqsubseteq 
        (\acarg, \auarg_2, \stenv_2, \henv_2)$ \quad iff \quad
        $\auarg_1 \sqsubseteq \auarg_2 \;\land\; \stenv_1 \sqsubseteq \stenv_2 
        \;\land\; \henv_1 \sqsubseteq \henv_2$
        \\
        $\henv_1 \sqsubseteq \henv_2$
        \quad iff \quad
        $\henv_1(u) \sqsubseteq \henv_2(u)$\quad for each $u \in \dom{\henv_1}$
        \\
        $\tfenv_1::\stenv_1 \sqsubseteq \tfenv_2::\stenv_2$
        \quad iff \quad
        $\tfenv_1 \sqsubseteq \tfenv_2 \;\land\; \stenv_1 \sqsubseteq \stenv_2$
        \\
        $\tuple{} \sqsubseteq \tuple{}$
        \\
        $\tfenv_1 \sqsubseteq \tfenv_2$
        \quad iff \quad
        $\tfenv_1(v) \sqsubseteq \tfenv_2(v)$\quad for each $v\in\dom{\tfenv_1}$
        \\
        $\auarg_1 \sqsubseteq \auarg_2$ \quad iff \quad 
        $\auarg_1 \subseteq \auarg_2$
        \\
        $\acarg \sqsubseteq \acarg$
      }}}
  \caption{The $\sqsubseteq$ relation on abstract states\label{fig:less-than}}
\end{figure}

We proceed to show that the \cfat{} semantics safely approximates the concrete
semantics.
First, we define a map \ctoa{\cdot} from concrete to abstract states.
Next, we show that if \cstat{} transitions to $\cstat'$ in the concrete
semantics, the abstract counterpart \ctoa{\cstat} of \cstat{} transitions to
a state $\astat'$ which approximates \ctoa{\cstat'}.
Hence, we ensure that the possible behaviors of the abstract interpreter include
the actual run-time behavior of the program.

The map \ctoa{\cdot} appears in Fig.~\ref{fig:conc-to-abs}.
The abstraction of an \deval{} state \cstat{} of the form
$(\denot{\slp g_1 \dots g_n\srp^{\psi}}, \cbenv, \cvenv, t)$
is an \daeval{} state \astat{} with the same call site.
Since \cstat{} does not have a stack, we must expose stack-related information 
hidden in \cbenv{} and \cvenv.
Assume that \ilam{l} is the innermost user lambda that contains $\psi$.
To reach $\psi$, control passed from a \dauapply{} state $\astat'$ 
over \ilam{l}.
According to our stack policy, the top frame contains bindings for the 
formals of \ilam{l} and any temporaries added along the path from $\astat'$ 
to \astat.
Therefore, the domain of the top frame is a subset of \ltov{l}, \ie{},
a subset of \ltov{\psi}.
For each user variable $u_i \in (\ltov{\psi} \cap \dom{\cbenv})$,
the top frame contains $\onemap{u_i}{\ctoa{\cvenv(u_i, \cbenv(u_i))}}$.
Let $k$ be the sole continuation variable in \ltov{\psi}.
If $\cvenv(k, \cbenv(k))$ is \haltcont{} (the return continuation is the 
top-level continuation), the rest of the stack is empty.
If $\cvenv(k, \cbenv(k))$ is $(\denot{\clam}, \cbenv')$, the second frame is for
the user lambda in which \ilam{\gamma} was born, and so forth:
proceeding through the stack, we add a frame for each live activation of a 
user lambda until we reach \haltcont.

The abstraction of a \duapply{} state over \tuple{\denot{\ulam}, \cbenv} 
is a \dauapply{} state \astat{} whose operator is \denot{\ulam}.
The stack of \astat{} is the stack in which the continuation 
argument was created, and we compute it using \tostackNA{} as above.

Abstracting a \dcapply{} is similar to the \duapply{} case,
only now the top frame is the environment of the continuation operator.
Note that the abstraction maps drop the time of the concrete states,
since the abstract states do not use times.

The abstraction of a user closure is the singleton set with the 
corresponding lambda.
The abstraction of a continuation closure is the corresponding lambda.
When abstracting a variable environment \cvenv, we only keep heap variables.

We can now state our simulation theorem.
The proof proceeds by case analysis on the concrete transition relation.
The relation $\astato \sqsubseteq \astatw$ is a partial order on abstract 
states and can be read as ``\astato{} is more precise than \astatw'' 
(Fig.~\ref{fig:less-than}).
The proof can be found in the appendix.
\begin{thm}[Simulation] 
If $\cstat \cstep \cstat'$ and $\ctoa{\cstat} \sqsubseteq \astat$, then 
there exists $\astat'$ such that $\astat \astep \astat'$
and $\ctoa{\cstat'} \sqsubseteq \astat'$.
\end{thm}

\section{Computing \cfat\label{sec:summarization}}

\subsection{Pushdown models and summarization}

In section \ref{sec:kcfa}, we saw that a monovariant analysis like \cfa{0}
treats the control-flow graph of \tw{len} as a finite-state machine (FSM),
where all paths are valid executions.
For $k>0$, \kcfa{} still approximates \tw{len} as a FSM, albeit a larger one
that has several copies of each procedure, caused by different call strings.

But in reality, calls and returns match; the call from 2 returns to 3 and each
call from 10 returns to 11.
Thus, by thinking of executions as strings in a context-free language, we can do
more precise flow analysis.
We can achieve this by approximating \tw{len} as a pushdown system (PDS)
\cite{journal/entcs/97/finkel/pds, conf/concur/97/bouajjani/pds}.
A PDS is similar to a pushdown automaton, except it does not read input from a
tape.
For illustration purposes, we take the (slightly simplified) view that the state
of a PDS is a pair of a program point and a stack.
The transition rules for call nodes push the return point on the stack:
$$(2, s) \pdsstep (5, 3::s), \qquad (10, s) \pdsstep (5, 11::s)$$
Function exits pop the node at the top of the stack and jump to it:
$$(13, n::s) \pdsstep (n, s)$$
All other nodes transition to their successor(s) and leave the stack unchanged,
\eg
$$(3,s) \pdsstep (4,s),\qquad (7,s) \pdsstep (8,s),\qquad (7,s) \pdsstep (9,s)$$

Suppose we want to find all nodes reachable from 1.
Obviously, we cannot do it by enumerating all states.
Thus, algorithms for pushdown reachability use a dynamic programming technique
called \emph{summarization}.
The intuition behind summarization is to flow facts from a program point $n$ 
with an \emph{empty} stack to a point $n'$ in the same procedure.
We say that $n'$ is \emph{same-context reachable} from $n$.
These facts are then suitably combined to get flow facts for the whole program.

We use summarization to explore the state space in CFA2.
Our algorithm is based on Sharir and Pnueli's functional approach 
\cite[pg.\ 207]{book/flowanalysis/81/sharir/interproc}, adapted to the more
modern terminology of Reps \etal~\cite{conf/popl/95/reps/interproc}.
Summarization requires that we know all call sites of a function.
Therefore, it does not apply directly to higher-order languages, because we
cannot find the call sites of a function by looking at a program's source code.
We need a \emph{search-based} variant of summarization, which records callers as
it discovers them.

We illustrate our variant on \tw{len}.
We find reachable nodes by recording \emph{path edges}, \ie, edges whose source
is the entry of a procedure and target is some program point in the same 
procedure.
Path edges should not be confused with the edges already present in \tw{len}'s
control-flow graph.
They are artificial edges used by the analysis to represent intraprocedural 
paths, hence the name.
From 1 we can go to 2, so we record \tuple{1,1} and \tuple{1,2}.
Then 2 calls \tw{len}, so we record the call \tuple{2, 5} and jump to 5.
In \tw{len}, we reach 6 and 7 and record \tuple{5,5}, \tuple{5,6} and 
\tuple{5,7}.
We do not assume anything about the result of the test, so we must follow both
branches.
By following the false branch, we discover \tuple{5,8} and \tuple{5,13}.
Node 13 is an exit, so each caller of \tw{len} can reach its corresponding 
return point.
We keep track of this fact by recording the \emph{summary} edge \tuple{5,13}.
We have only seen a call from 2, so we return to 3 and record \tuple{1,3}.
Finally, we record \tuple{1,4}, which is the end of the program.
By analyzing the true branch, we discover edges \tuple{5,9} and \tuple{5,10},
and record the new call \tuple{10,5}.
Reachability inside \tw{len} does not depend on its calling context, so from the
summary edge \tuple{5,13} we infer that 10 can reach 11 and we record 
\tuple{5,11} and subsequently \tuple{5,12}.
At this point, we have discovered all possible path edges.

Summarization works because we can temporarily forget the caller while
analyzing inside a procedure, and remember it when we are about to return.
A consequence is that if from node $n$ with an empty stack we can reach $n'$ 
with stack $s$, then $n$ with $s'$ can go to $n'$ with $\ti{append}(s, s')$.

\subsection{Local semantics\label{subsec:localsems}}

\begin{figure}[!t]
  {\small
    \begin{tabular}{@{} l @{$\qquad\quad$} l @{}}
      \lbox{
        \raisebox{0.7cm}{
          $\quad\aubiga{e, \psi, \tfenv, \henv} \triangleq
          \begin{cases}
            \mset{e} & \islam{e} \\
            \tfenv(e) & \instack{\psi, e} \\
            \henv(e) & \inheap{\psi, e}
          \end{cases}$
        }
        \\ \\
        \lablr{UEA}\;
        \lbox{
          $(\denot{\ucall}, \tfenv, \henv) \lstep (\mulam, \auarg, \henv)$ \\
          $\mulam \in \aubiga{f, l, \tfenv, \henv}$ \\
          $\auarg = \aubiga{e, l, \tfenv, \henv}$
        } \\ \\
        \lablr{UAE}\;
        \lbox{
          $(\denot{\ulam}, \auarg, \henv) \lstep 
          (\mcall, \onemap{u}{\auarg}, \henv')$ \\
          $\henv' = 
          \begin{cases}
            \henv\join\onemap{u}{\auarg} & \inheap{u} \\
            \henv & \instack{u}
          \end{cases} $
        } \\ \\
        \lablr{CEA}\;
        \lbox{
          $(\denot{\slp \mclam \; e\srp^{\gamma}}, \tfenv, \henv) \lstep 
          (\mclam, \auarg, \tfenv, \henv)$ \\
          $\auarg = \aubiga{e, \gamma, \tfenv, \henv}$ 
        } \\ \\
        \lablr{CAE}\;
        \lbox{
          $(\denot{\clam}, \auarg, \tfenv, \henv) \lstep
          (\mcall, \tfenv', \henv')$ \\
          $\tfenv' = \tfenv\onemap{u}{\auarg}$ \\
          $\henv' = 
          \begin{cases}
            \henv\join\onemap{u}{\auarg} & \inheap{u} \\
            \henv & \instack{u}
          \end{cases}$
        }
      }
      &
      \lbox{
        ~ \\
        \begin{tabular}{@{} r @{\;\;} c @{\;} l @{}}
          \multicolumn{3}{c}{Local domains:\makebox[3cm][c]{}} 
          \rule{0cm}{0.8cm} \\
          $\dleval$ & $=$ & $\dcall \times \dlstack \times \dheap$ 
          \rule{0cm}{0.42cm}\\
          $\dluapply$ & $=$ & $\dulam \times \dauclos \times \dheap$
          \rule{0cm}{0.42cm}\\
          $\dlcapply$ & $=$ & 
          \lbox{
            $\dacclos \times \dauclos \times \dlstack \times$ 
            $ \dheap$}
          \rule{0cm}{0.42cm}\\
          $\dlframe$ & $=$ & $\duvar \rightharpoonup \dauclos$
          \rule{0cm}{0.42cm}\\
          $\dlstack$ & $=$ & $\dlframe$
          \rule{0cm}{0.42cm} 
        \end{tabular}
        \\ \\
        \begin{tabular}{@{$\quad$} l @{}}
          Abstract to local maps: \rule{0cm}{0.6cm} \\
          $\atol{(\mcall, \stenv, \henv)} = (\mcall, \atol{\stenv}, \henv)$ 
          \rule{0cm}{0.42cm}\\
          $\atol{(\mulam, \auarg, \acarg, \stenv, \henv)} = (\mulam, \auarg, \henv)$
          \rule{0cm}{0.42cm}\\
          $\atol{(\acarg, \auarg, \stenv, \henv)} = (\acarg, \auarg, \atol{\stenv}, \henv)$
          \rule{0cm}{0.42cm} \\
          $\atol{\tfenv::\stenv'} = \msetcomp{(u, \tfenv(u)) \!}{\!\duvar_?(u)}$ 
          \rule{0cm}{0.42cm}\\
          $\atol{\tuple{}} = \emptyset$ 
          \rule{0cm}{0.42cm}
        \end{tabular}
      }
    \end{tabular}
  }
  \caption{Local semantics \label{fig:localsems}}
\end{figure}

Summarization-based algorithms operate on a finite set of program points.
Hence, we cannot use (an infinite number of) abstract states as program points.
For this reason, we introduce \emph{local states} and define 
a map \atol{\cdot} from abstract to local states (Fig.~\ref{fig:localsems}).
Intuitively, a local state is like an abstract state but with a single frame
instead of a stack.
Discarding the rest of the stack makes the local state space finite;
keeping the top frame allows precise lookups for stack references.

The local semantics describes executions that do not touch the rest of the stack
(in other words, executions where functions do not return).
Thus, a \dlceval{} state with call site \kcall{} has no successor in this 
semantics.
Since functions do not call their continuations, the frames of local states 
contain only user bindings.
Local steps are otherwise similar to abstract steps.
The metavariable \lstat{} ranges over local states.
We define the map \ctol{\cdot} from concrete to local states to be
$\atol{\cdot} \circ \ctoa{\cdot}$.

We can now see how the local semantics fits in a summarization algorithm for
\cfat.
Essentially, \cfat{} approximates a higher-order program as a PDS.
The local semantics describes the PDS transitions that do not return
(intraprocedural steps and function calls).
We discover return points by recording callers and summary edges.

Summarization distinguishes between different kinds of states: entries, exits, 
calls, returns and inner states.
\cps{} lends itself naturally to such a categorization:
\begin{enumerate}[$\bullet$]
\item
  A \dluapply{} state corresponds to an \tb{entry} node---control is about to 
  enter the body of a function.
\item
  A \dlceval{} state where the operator is a variable is an \tb{exit} 
  node---a function is about to pass its result to its context.
\item
  A \dlceval{} state where the operator is a lambda is an \tb{inner} state.
\item
  A \dlueval{} state where the continuation argument is a variable 
  is an \tb{exit}---at tail calls control does not return to the caller.
\item
  A \dlueval{} state where the continuation argument is a lambda is a \tb{call}.
\item
  A \dlcapply{} state is a \tb{return} if its predecessor is an exit, 
  or an \tb{inner} state if its predecessor is also an inner state.
  Our algorithm will not need to distinguish between the two kinds of
  \dlcapply{}s; the difference is just conceptual.
\end{enumerate}\medskip

\noindent Last, we generalize the notion of summary edges to handle tail recursion.
Consider an earlier example, written in \cps.
\begin{center}
  \begin{minipage}{0.75\columnwidth}
    \begin{alltt}
((\lam(app id k) 
   (app id 1 (\ilam{1}(n1) (app id 2 (\ilam{2}(n2) (+ n1 n2 k))))))
 (\lam(f e k) (f e k))
 (\lam(x k) (k x))
 \haltcont) \end{alltt}    
  \end{minipage}
\end{center}
The call \tw{(f e k)} in the body of \tw{app} is a tail call, so no continuation
is born there.
Upon return from the first call to \tw{id}, we must be careful to pass the 
result to \ilam{1}.
Also, we must restore the environment of the first call to \tw{app}, \emph{not}
the environment of the tail call.
Sim\-i\-lar\-ly, the second call to \tw{id} must return to \ilam{2} and restore
the correct environment.
We achieve these by recording a ``cross-procedure'' summary from the entry of 
\tw{app} to call site \tw{(k x)}, which is the exit of \tw{id}.
This transitive nature of summaries is essential for tail recursion.

\subsection{Summarization for \cfat \label{subsec:summarization}}

\begin{figure}[!t]
\newcommand{\tmphack}
           {\eh{\begin{cases}
                 \tfenv_2\onemap{f}{\mset{\denot{\slp\lambda_{l_3}\slp\mbox{}u_3 \; k_3\srp \;\mcall_3\srp}}} & \instack{l_2, f} \\ 
                 \tfenv_2 & \inheap{l_2, f} \lor \islam{f} 
           \end{cases}}}
{\footnotesize
\begin{alltt}
          01    \summary, \callers, \tcallers, \finals \assgn \(\emptyset\)
          02    \seen, \work \assgn \(\mset{(\linitstate, \linitstate)}\)
          03    while \work \(\neq \emptyset\)
          04      remove (\lstato, \lstatw) from \work
          05      switch \lstatw
          06        case \lstatw of Entry, CApply, Inner-CEval
          07          for each \lstath in \succ{\lstatw} Propagate(\lstato, \lstath)
          08        case \lstatw of Call
          09          for each \lstath in \succ{\lstatw}
          10            Propagate(\lstath, \lstath)
          11            insert (\lstato, \lstatw, \lstath) in \callers
          12            for each (\lstath, \lstatf) in \summary  Update(\lstato, \lstatw, \lstath, \lstatf)
          13        case \lstatw of Exit-CEval
          14          if \lstato = \linitstate then
          15            Final(\lstatw)
          16          else
          17            insert (\lstato, \lstatw) in \summary
          18            for each (\lstath, \lstatf, \lstato) in \callers  Update(\lstath, \lstatf, \lstato, \lstatw)
          19            for each (\lstath, \lstatf, \lstato) in \tcallers Propagate(\lstath, \lstatw)
          20        case \lstatw of Exit-TC
          21          for each \lstath in \succ{\lstatw}
          22            Propagate(\lstath, \lstath)
          23            insert (\lstato, \lstatw, \lstath) in \tcallers
          24            for each (\lstath, \lstatf) in \summary Propagate(\lstato, \lstatf)
                Propagate(\lstato, \lstatw) \(\triangleq\) \rule{0cm}{0.45cm}
          25      if (\lstato, \lstatw) not in \seen then insert (\lstato, \lstatw) in \seen and \work
                Update(\lstato, \lstatw, \lstath, \lstatf) \(\triangleq\)  \rule{0cm}{0.45cm}
          26      \lstato of the form  \((\denot{\slp\lambda\sb{l\sb{1}}\slp\mbox{}u\sb{1} k\sb{1}\srp \mcall\sb{1}\srp} , \auarg\sb{1}, \henv\sb{1})\)
          27      \lstatw of the form  \((\denot{\slp\mbox{}f e\sb{2} \slp\lambda\sb{\gamma\sb{2}} \slp\mbox{}u\sb{2}\srp \mcall\sb{2}\srp\srp\sp{l\sb{2}}}, \tfenv\sb{2}, \henv\sb{2})\)
          28      \lstath of the form  \((\denot{\slp\lambda\sb{l\sb{3}}\slp\mbox{}u\sb{3} k\sb{3}\srp \mcall\sb{3}\srp} , \auarg\sb{3}, \henv\sb{2})\)
          29      \lstatf of the form  \((\denot{\slp\mbox{}k\sb{4} e\sb{4}\srp\sp{\gamma\sb{4}}}, \tfenv\sb{4}, \henv\sb{4})\)
          30      \auarg \assgn  \aubiga{e\sb{4}, \gamma\sb{4}, \tfenv\sb{4}, \henv\sb{4}}
          31      \tfenv \assgn  \tmphack
          32      \lstat \assgn  \((\denot{\slp\mbox{}\lambda\sb{\gamma\sb{2}}\slp\mbox{}u\sb{2}\srp \mcall\sb{2}\srp}, \auarg, \tfenv, \henv\sb{4})\)
          33      Propagate(\lstato, \lstat)  
                Final(\lstat) \(\triangleq\)  \rule{0cm}{0.45cm}
          34      \lstat of the form  \((\denot{\kcall}, \tfenv, \henv)\)
          35      insert  \((\haltcont, \aubiga{e, \gamma, \tfenv, \henv}, \emptyset, \henv)\) in \finals 
\end{alltt}
}
\caption{\cfat{} workset algorithm\label{fig:workset}}
\end{figure}

The algorithm for \cfat{} is shown in Fig.~\ref{fig:workset}.
It is a search-based summarization for higher-order programs with tail calls.
Its goal is to compute which local states are reachable from the initial 
state of a program through paths that respect call/return matching.

\paragraph{\myp{Overview of the algorithm's structure}}
The algorithm uses a workset \work, which contains path edges and summaries to
be examined.
An edge $(\lstato, \lstatw)$ is an ordered pair of local states.
We call \lstato{} the \emph{source} and \lstatw{} the \emph{target} of the edge.
At every iteration, we remove an edge from \work{} and process it, potentially 
adding new edges in \work.
We stop when \work{} is empty.

The algorithm maintains several sets.
The results of the analysis are stored in the set \seen.
It contains path edges (from a procedure entry to a state in the same procedure)
and summary edges (from an entry to a \dlceval{} exit, not necessarily in the
same procedure).
The target of an edge in \seen{} is reachable from the source and from the
initial state (\confer{} the\-o\-rem \ref{thm:sum/tion-sound}).
Summaries are also stored in \summary.
\finals{} records final states, \ie, \dlcapply s that call \haltcont{} with a 
return value for the whole program.
\callers{} contains triples \tuple{\lstato, \lstatw, \lstath}, where \lstato{}
is an entry, \lstatw{} is a call in the same procedure and \lstath{} is the
entry of the callee.
\tcallers{} contains triples \tuple{\lstato, \lstatw, \lstath}, where \lstato{}
is an entry, \lstatw{} is a tail call in the same pro\-ce\-dure and \lstath{} 
is the entry of the callee.
The initial state \linitstate{} is defined as \ctol{\initstate}.
The helper function \succ{\lstat} returns the successor(s) of \lstat{} according
to the local semantics.

\paragraph{\myp{Edge processing}}
Each edge $(\lstato, \lstatw)$ is processed in one of four ways, depending on 
\lstatw.
If \lstatw{} is an entry, a return or an inner state (line 6), 
then its successor \lstath{} is a state in the same procedure.
Since \lstatw{} is reachable from \lstato, 
\lstath{} is also reachable from \lstato.
If we have not already recorded the edge $(\lstato, \lstath)$, we do it now 
(line 25).

If \lstatw{} is a call (line 8) then \lstath{} is the entry of the callee,
so we propagate $(\lstath, \lstath)$ instead of $(\lstato, \lstath)$ (line 10).
Also, we record the call in \callers.
If an exit \lstatf{} is reachable from \lstath, it should return
to the continuation born at \lstatw{} (line 12).
The function \tw{Update} is responsible for computing the return state.
We find the return value \auarg{} by evaluating the expression $e_4$ passed to
the continuation (lines 29-30).
Since we are returning to \ilam{\gamma_2}, we must restore the environment of 
its creation which is $\tfenv_2$ (possibly with stack filtering, line 31).
The new state \lstat{} is the corresponding return of \lstatw{},
so we propagate $(\lstato, \lstat)$ (lines 32-33).

If \lstatw{} is a \dlceval{} exit and \lstato{} is the initial state (lines 
14-15), then \lstatw{}'s successor is a final state (lines 34-35).
If \lstato{} is some other entry, we record the edge in \summary{} and pass the
re\-sult of \lstatw{} to the callers of \lstato{} (lines 17-18).
Last, consider the case of a tail call \lstatf{} to \lstato{} (line 19).
No continuation is born at \lstatf.
Thus, we must find where \lstath{} (the entry that led to the tail call) was 
called from.
Then again, all calls to \lstath{} may be tail calls, in which case we keep 
searching further back in the call chain to find a return point.
We do the backward search by transitively adding a cross-procedure summary from
\lstath{} to \lstatw{} (line 25).

If \lstatw{} is a tail call (line 20), we find its successors and record the 
call in \tcallers{} (lines 21-23).
If a successor of \lstatw{} goes to an exit, we propagate a cross-procedure
summary transitively (line 24).
Figure \ref{fig:cfa2-eg} shows a complete run of the algorithm for a small 
program.

\begin{figure}[!t]
\newcommand{\calli}{\tw{(k x)}}
\newcommand{\lami}{\tw{(\ilam{1}(x k)\calli)}}
\newcommand{\calliii}{\tw{(id 2 h)}}
\newcommand{\lamiii}{\tw{(\ilam{3}(u)\calliii)}}
\newcommand{\callii}{\tw{(id 1 \lamiii)}}
\newcommand{\lamii}{\tw{(\ilam{2}(id h)\callii)}}
\newcommand{\statnum}[1]{\eh{\tilde{\varsigma}_{#1}}}
\newcommand{\mtset}{\eh{\emptyset}}
{\small
\begin{tabular*}{\textwidth}{c}
  {\renewcommand{\arraystretch}{1.4}\!\!
    \begin{tabular}{| l | l | l |}
      \hline
      Name & Kind & Value \\
      \hline
      \linitstate & Entry & $(\denot{\lamii}, \mset{\denot{\lami}}, \mtset)$ \\
      \hline
      \statnum{1} & Call & 
      $(\denot{\callii}, \onemap{\tw{id}}{\mset{\ilam{1}}}, \mtset)$ \\
      \hline
      \statnum{2} & Entry & $(\ilam{1}, \mset{1}, \mtset)$ \\
      \hline
      \statnum{3} & Exit \dlceval & 
      $(\denot{\calli}, \onemap{\tw{x}}{\mset{1}}, \mtset)$ \\
      \hline
      \statnum{4} & \dlcapply &
      $(\ilam{3}, \mset{1}, \onemap{\tw{id}}{\mset{\ilam{1}}}, \mtset)$ \\
      \hline
      \statnum{5} & Exit tail call & $(\denot{\calliii}, 
      \onemap{\tw{id}}{\mset{\ilam{1}}}\onemap{\tw{u}}{\mset{1}}, \mtset)$ \\
      \hline
      \statnum{6} & Entry & $(\ilam{1}, \mset{2}, \mtset)$ \\
      \hline
      \statnum{7} & Exit \dlceval & 
      $(\denot{\calli}, \onemap{\tw{x}}{\mset{2}}, \mtset)$ \\
      \hline
      \statnum{8} & \dlcapply{} (final state) & 
      $(\haltcont, \mset{2}, \mtset, \mtset)$ \\
      \hline
    \end{tabular}
  }
  \\ \\
  \begin{tabular}{|@{$\quad$} c @{$\quad$} | @{$\quad$} c @{$\quad$} | @{$\quad$} c @{$\quad$} | @{$\quad$} c @{$\quad$} | @{$\qquad$} c @{$\qquad$}|}
    \hline
    \work    &    \summary    &      \callers     &   \tcallers   &   \finals \\
    \hline 
    $(\linitstate, \linitstate)$ & 
    \mtset & \mtset & \mtset & \mtset \rule{0cm}{0.45cm} 
    \\
    \hline
    $(\linitstate, \statnum{1})$ &
    \mtset & \mtset & \mtset & \mtset \rule{0cm}{0.45cm} 
    \\
    \hline
    $(\statnum{2}, \statnum{2})$ & \mtset & 
    $(\linitstate, \statnum{1}, \statnum{2})$ & 
    \mtset & \mtset \rule{0cm}{0.45cm} 
    \\
    \hline
    $(\statnum{2}, \statnum{3})$ & \mtset & 
    $(\linitstate, \statnum{1}, \statnum{2})$ &
    \mtset & \mtset \rule{0cm}{0.45cm} 
    \\
    \hline
    $(\linitstate, \statnum{4})$ &
    $(\statnum{2}, \statnum{3})$ & 
    $(\linitstate, \statnum{1}, \statnum{2})$ &
    \mtset & \mtset \rule{0cm}{0.45cm} 
    \\
    \hline
    $(\linitstate, \statnum{5})$ &
    $(\statnum{2}, \statnum{3})$ & 
    $(\linitstate, \statnum{1}, \statnum{2})$ &
    \mtset & \mtset \rule{0cm}{0.45cm} 
    \\
    \hline
    $(\statnum{6}, \statnum{6})$ &
    $(\statnum{2}, \statnum{3})$ & 
    $(\linitstate, \statnum{1}, \statnum{2})$ &
    $(\linitstate, \statnum{5}, \statnum{6})$ &
    \mtset \rule{0cm}{0.45cm} 
    \\
    \hline
    $(\statnum{6}, \statnum{7})$ &
    $(\statnum{2}, \statnum{3})$ & 
    $(\linitstate, \statnum{1}, \statnum{2})$ &
    $(\linitstate, \statnum{5}, \statnum{6})$ &
    \mtset \rule{0cm}{0.45cm} 
    \\
    \hline
    $(\linitstate, \statnum{7})$ &
    $(\statnum{2}, \statnum{3}), \; (\statnum{6}, \statnum{7})$ & 
    $(\linitstate, \statnum{1}, \statnum{2})$ &
    $(\linitstate, \statnum{5}, \statnum{6})$ &
    \mtset \rule{0cm}{0.45cm} 
    \\
    \hline
    \mtset &
    $(\statnum{2}, \statnum{3}), \; (\statnum{6}, \statnum{7})$ & 
    $(\linitstate, \statnum{1}, \statnum{2})$ &
    $(\linitstate, \statnum{5}, \statnum{6})$ &
    \statnum{8} \rule{0cm}{0.45cm} 
    \\
    \hline
  \end{tabular}
\end{tabular*}
}
\caption{A complete run of \cfat{}. 
  Note that \ilam{1} is applied twice and returns to the correct context both 
  times. 
  The program evaluates to 2.
  For brevity, we first show all reachable states and then refer to them by 
  their names.
  \linitstate{} shows the whole program; in the other states we abbreviate
  lambdas by their labels.
  All heaps are $\emptyset$ because there are no heap variables.
  The rows of the table show the contents of the sets at line 3 for each
  iteration.
  \seen{} contains all pairs entered in \work{}.
  \label{fig:cfa2-eg}}
\end{figure}

\subsection{Correctness of the workset algorithm}

The local state space is finite, so there is a finite number of path and 
summary edges.
We record edges as seen when we insert them in \work{},
which ensures that no edge is inserted in \work{} twice.
Therefore, the algorithm terminates.

We obviously cannot visit an infinite number of abstract states.
To establish the soundness of our analysis, we show that if a state \astat{} is
reachable from \ainitstate, then the algorithm visits \atol{\astat} 
(\confer{} theorem \ref{thm:sum/tion-sound}). 
For instance, \cfat{} on \tw{len} tells us that we reach program point 5, 
\emph{not} that we reach 5 with a stack of size 1, 2, 3, \etc.

Soundness guarantees that \cfat{} does not miss any flows,
but it may also add flows that do not happen in the abstract semantics.
For example, a sound but useless algorithm would add all pairs of local states 
in \seen{}.
We establish the completeness of \cfat{} by proving that every visited
edge corresponds to an abstract flow (\confer{} theorem 
\ref{thm:sum/tion-complete}), which means that there is no loss in precision 
when going from abstract to local states.

The theorems use two definitions.
The first associates a state \astat{} with its \emph{corresponding entry},
\ie{}, the entry of the procedure that contains \astat.
The second finds all entries that reach \coren{\astat} through tail calls.
We include the proofs of the theorems in the appendix.
\begin{defi}
  The Corresponding Entry \coren{\astat} of a state {\astat} in a path $p$ is:
  \begin{enumerate}[$\bullet$]
  \item
    \astat, if {\astat} is an Entry
  \item
    \astato, if {\astat} is not an Entry, \astatw{} is not an Exit-CEval,
    $p \equiv p_1 \astep \astato \asteps \astatw \astep \astat \astep p_2$,
    and $\coren{\astatw} = \astato$
  \item
    \astato, if {\astat} is not an Entry, 
    $\:p \equiv p_1 \astep \astato \astepso \astatw \astep \astath
    \astepso \astatf \astep \astat \astep p_2$, \astatw{} is a Call 
    and \astatf{} is an Exit-CEval, $\coren{\astatw} = \astato$, and 
    $\astath \in \corens{\astatf}$
  \end{enumerate}
\end{defi}

\begin{defi}
  For a state {\astat} and a path $p$, \corens{\astat} is the
  smallest set such that: 
  \begin{enumerate}[$\bullet$]
  \item
    $\coren{\astat} \in \corens{\astat}$
  \item
    $\corens{\astato} \subseteq \corens{\astat}$,
    when $p \equiv p_1 \astep \astato \astep \astatw \asteps \astat 
    \astep p_2$, \astato{} is a Tail Call,
    \astatw{} is an Entry, and $\astatw = \coren{\astat}$
  \end{enumerate}
\end{defi}
\begin{thm}[Soundness]\label{thm:sum/tion-sound} 
  If $p \equiv \ainitstate \asteps \astat$ then, after summarization:
  \begin{enumerate}[$\bullet$]
  \item
    if $\astat$ is not a final state then
    $(\atol{\coren{\astat}}, \atol{\astat}) \in \seen$
  \item
    if $\astat$ is a final state then $\atol{\astat} \in \finals$
  \item
    if $\astat$ is an Exit-CEval and $\astat' \!\in\! \corens{\astat}$
    then $(\atol{\astat'}, \atol{\astat}) \!\in\! \seen$
  \end{enumerate}
\end{thm}
\begin{thm}[Completeness]\label{thm:sum/tion-complete} 
  After summarization:
  \begin{enumerate}[$\bullet$]
  \item
    For each $(\lstato, \lstatw)$ in \seen,
    there exist \astato, \astatw{} and $p$ such that 
    $p \equiv \ainitstate \asteps \astato \asteps \astatw$ \;and\; 
    $\lstato = \atol{\astato}$ \;and\;
    $\lstatw = \atol{\astatw}$ \;and\;
    $\astato \in \corens{\astatw}$
  \item
    For each \lstat{} in \finals, 
    there exist \astat{} and $p$ such that 
    $p \equiv \ainitstate \astepso \astat$ \;and\;
    $\lstat = \atol{\astat}$ \;and\;
    \astat{} is a final state.
  \end{enumerate}
\end{thm}

\subsection{Complexity\label{subsec:complexity}}

A simple calculation shows that \cfat{} is in \textsc{exptime}.
The size of the domain of \dheap{} is $n$ and the size of the range is $2^n$,
so there are $2^{n^2}$ heaps.
Similarly, there are $2^{n^2}$ frames.
The size of \dlstate{} is dominated by the size of \dlcapply{} which is
$n \cdot 2^n \cdot 2^{n^2} \cdot 2^{n^2} = O(n \cdot 2^{2n^2 + n})$.
The size of \seen{} is the product of the sizes of \dluapply{} and \dlstate{},
which is $(n \cdot 2^n \cdot 2^{n^2}) \cdot (n \cdot 2^{2n^2 + n}) = 
O(n^2 \cdot 2^{3n^2 + 2n})$.

The running time of the algorithm is bounded by the number of edges in \work{} 
times the cost of each iteration.
\work{} contains edges from \seen{} only, so its size is 
$O(n^2 \cdot 2^{3n^2 + 2n})$.
The most expensive iteration happens when line 19 is executed.
There are $O(n^3 \cdot 2^{4n^2 + 2n})$ \tcallers{} and for each one we call 
\tw{Propagate}, which involves searching \seen.
Therefore, the loop costs 
$O(n^3 \cdot 2^{4n^2 + 2n}) \cdot O(n^2 \cdot 2^{3n^2 + 2n}) = 
O(n^5 \cdot 2^{7n^2 + 4n})$.
Thus, the total cost of the algorithm is 
$O(n^2 \cdot 2^{3n^2 + 2n}) \cdot O(n^5 \cdot 2^{7n^2 + 4n}) = 
 O(n^7 \cdot 2^{10n^2 + 6n})$.

Showing that \cfat{} is in \textsc{exptime} does not guarantee the existence of
a program that, when analyzed, triggers the exponential behavior. 
Is there a such a program? The answer is yes.
Consider the following program, suggested to us by Danny Dub\'e:
{\footnotesize
\begin{alltt}
                        (let* ((merger   (\ilam{1}(f) (\ilam{2}(_) f)))
                               (_        (merger (\ilam{3}(x) x)))
                               (clos     (merger (\ilam{4}(y) y)))
                               (f1       (clos _)\(\sb{1}\))
                               (_        (f1 _)\(\sb{1'}\))
                               (f2       (clos _)\(\sb{2}\))
                               (_        (f2 _)\(\sb{2'}\))
                                       \(\vdots\)
                               (fn       (clos _)\(\sb{n}\))
                               (_        (fn _)\(\sb{n'}\)))
                          _ ) \end{alltt}}
The idea is to create an exponential number of frames by exploiting the strong
updates \cfat{} does on the top frame.
The code is in direct style for brevity;
the let-bound variables would be bound by continuation lambdas in the equivalent
\cps{} program. 
The only heap reference appears in the body of \ilam{2}.
We use underscores for unimportant expressions.

The \tw{merger} takes a function, binds \tw{f} to it and returns a closure that
ignores its argument and returns \tw{f}.
We call the \tw{merger} twice so that \tw{f} is bound to 
\mset{\ilam{3}, \ilam{4}} in the heap.
Now \tw{clos} is bound to \ilam{2} in the top frame and every call to \tw{clos}
returns \mset{\ilam{3}, \ilam{4}}.
Thus, after call site 1 the variable \tw{f1} is bound to 
\mset{\ilam{3}, \ilam{4}}.
At $1'$, execution splits in two branches.
One calls \ilam{3} and filters the binding of \tw{f1} in the top frame to
\mset{\ilam{3}}.
The other calls \ilam{4} and filters the binding to \mset{\ilam{4}}.
Each branch will split in two more branches at call $2'$, \etc.
By binding each \tw{fi} to a set of two elements and applying it immediately, we
force a strong update and create exponentially many frames.

Even though strong update can be subverted, it can also speed up the analysis
of some programs by avoiding spurious flows. 
In \tw{compose-same} (\confer{} sec.~\ref{sec:fake-rebind}), if two
lambdas \ilam{1} and \ilam{2} flow to \tw{f}, \cfa{0} will apply each lambda
at each call site, resulting in four flows.
\cfat{} will only examine two flows, one that uses \ilam{1} in both call sites
and one that uses \ilam{2}.

We tried to keep the algorithm of Fig.~\ref{fig:workset} simple because it is
meant to be a model.
There are many parameters one can tune to improve the performance and/or 
asymptotic complexity of \cfat:
\begin{enumerate}[$\bullet$]
\item 
  \emph{no stack filtering}: 
  \cfat{} is sound without stack filtering, but less precise.
  Permitting fake rebinding may not be too harmful in practice.
  Suppose that a set $\mset{\ilam{1}, \ilam{2}}$ flows to a variable $v$ with 
  two stack references $v_{l}$ and $v_{l'}$.
  Even with stack filtering, both lambdas will flow to both references.
  Stack filtering just prevents us from using \ilam{1} at $v_{l}$ and \ilam{2} 
  at $v_{l'}$ along the same execution path.
\item
  \emph{heap widening}: 
  implementations of flow analyses rarely use one heap per state. 
  They use a global heap instead and states carry timestamps 
  \cite[ch.\ 5]{diss/cmu/91/olin}.
  \dheap{} is a lattice of height $O(n^2)$.
  Since the global heap grows monotonically, it can change at most $O(n^2)$
  times during the analysis.
\item
  \emph{summary reuse}: we can avoid some reanalyzing of procedures by 
  creating general summaries that many callers can use. 
  One option is to create more approximate summaries by widening.
  Another option is to include only relevant parts of the state in the 
  summary~\cite{conf/pldi/09/chandra/snugglebug}.
\item
  \emph{representation of the sets}: in calculating the exponential upper 
  bound, we pessimistically assumed that looking up an element in a set takes
  time linear in the size of the set. 
  This need not be true if one uses efficient data structures to represent
  \seen{} and the other sets.
\end{enumerate}
An in-depth study of the performance and complexity of the proposed variants 
would increase our understanding of their relative merits.
Also, we do not know if \cfat{} has an exponential lower bound.
Our evaluation, presented in the next section, shows that \cfat{} compares 
favorably to \cfa{0}, a cubic algorithm.

\section{Evaluation\label{sec:evaluation}}

\noindent We implemented \cfat, \cfa{0} and \cfa{1} for the Twobit Scheme
compiler~\cite{conf/lfp/94/clinger/larceny}
and used them to do constant propagation and folding.
In this section we report on some initial measurements and comparisons.

\cfa{0} and \cfa{1} use a standard workset algorithm.
\cfat{} uses the algorithm of section~\ref{subsec:summarization}.
To speed up the analyses, the variable environment and the heap are global.

\begin{figure}[!t]
  \begin{tabular}{|l|r|r|r|c|r|c|r|c|}
    \hline
    ~           & ~   &  ~    & \multicolumn{2}{|c|}{\cfa{0}} & \multicolumn{2}{|c|}{\cfa{1}} & \multicolumn{2}{|c|}{\cfat} 
    \\ \cline{4-9}
    ~                  & $S_?$ & $H_?$ &  visited  & constants &   visited   & constants &   visited & constants
    \\ \hline
    \tw{len}                       &  9  &  0    &    81  &  0   &   126   &  0   &     55  &   2     
    \\ \hline
    \tw{rev-iter}                  & 17  &  0    &   121  &  0   &   198  &   0   &     82  &   4     
    \\ \hline
    \tw{len-Y}                     & 15  &  4    &   199  &  0   &   356  &   0   &    131  &   2     
    \\ \hline
    \tw{tree-count}                & 33  &  0    &   293  &  2   &  2856   &  6   &    183  &  10     
    \\ \hline
    \tw{ins-sort}                  & 33  &  5    &   509  &  0   &  1597   &  0   &    600  &   4     
    \\ \hline
    \tw{DFS}                       & 94  & 11    &  1337  &  8   &  6890   &  8   &   1719  &  16     
    \\ \hline
    \tw{flatten}                   & 37  &  0    &  1520  &  0   &  6865   &  0   &    478  &   5     
    \\ \hline
    \tw{sets}                      & 90  &  3    &  3915  &  0   & 54414   &  0   &   4251  &   4     
    \\ \hline
    \tw{church-nums}$\phantom{aaa}$& 46  & 23    & 19130  &  0   & 19411   &  0   &  22671  &   0   
    \\ \hline
  \end{tabular}
  \caption{Benchmark results \label{fig:benchmarks}}
\end{figure}

We compared the effectiveness of the analyses on a small set of 
benchmarks (Fig.~\ref{fig:benchmarks}).
We measured the number of stack and heap references in each program and the 
number of constants found by each analysis.
We also recorded  what goes in the workset in each analysis, \ie, the number of
abstract states visited by \cfa{0} and \cfa{1}, and the number of path and 
summary edges visited by \cfat.
The running time of an abstract interpretation is proportional to the amount of
things inserted in the workset.

We chose programs that exhibit a variety of control-flow patterns.
\tw{Len} computes the length of a list recursively. 
\tw{Rev-iter} reverses a list tail-recursively.
\tw{Len-Y} computes the length of a list using the Y-combinator instead of
explicit recursion.
\tw{Tree-count} counts the nodes in a binary tree.
\tw{Ins-sort} sorts a list of numbers using insertion-sort.
\tw{DFS} does depth-first search of a graph.
\tw{Flatten} turns arbitrarily nested lists into a flat list.
\tw{Sets} defines the basic set operations and tests De Morgan's laws on sets of
numbers.
\tw{Church-nums} tests distributivity of multiplication over addition for
a few Church numerals.

\cfat{} finds the most constants, followed by \cfa{1}.
\cfa{0} is the least precise.
\cfat{} is also more efficient at exploring its abstract state space.
In five out of nine cases, it visits fewer paths than \cfa{0} does states.
The visited set of \cfat{} can be up to 3.2 times smaller (\tw{flatten}), 
and up to 1.3 times larger (\tw{DFS}) than the visited set of \cfa{0}.
\cfa{1} is less efficient than both \cfa{0} (9/9 cases) and \cfat{} (8/9 cases).
The visited set of \cfa{1} can be significantly larger than that of \cfat{} in 
some cases (15.6 times in \tw{tree-count}, 14.4 times in \tw{flatten}, 
12.8 times in \tw{sets}).

Naturally, the number of stack references in a program is much higher than
the number of heap references; 
most of the time, a variable is referenced only by the lambda that binds it.
Thus, \cfat{} uses the precise stack lookups more often than the imprecise
heap lookups.

\section{Related work\label{sec:related}}

\noindent We were particularly influenced by Chaudhuri's paper on subcubic algorithms for
recursive state machines \cite{conf/popl/08/chaudhuri/subcubic}.
His clear and intuitive description of summarization helped us realize that we
can use this technique to explore the state space of \cfat.

Our workset algorithm is based on Sharir and Pnueli's functional approach 
\cite[pg.\ 207]{book/flowanalysis/81/sharir/interproc} and the tabulation 
algorithm of Reps \etal~\cite{conf/popl/95/reps/interproc}, extended for tail
recursion and higher-order functions.
In section \ref{subsec:localsems}, we mentioned that \cfat{} essentially 
produces a pushdown system.
Then, the reader may wonder why we designed a new algorithm instead of using an
existing one like $\mi{post}^*$~\cite{journal/entcs/97/finkel/pds, 
conf/concur/97/bouajjani/pds}.
The reason is that callers cannot be identified syntactically in higher-order
languages.
Hence, algorithms that analyze higher-order programs must be based on search.
The tabulation algorithm can be changed to use search fairly naturally.
It is less clear to us how to do that for $\mi{post}^*$.
In a way, \cfat{} creates a pushdown system and analyzes it \emph{at the same
time}, much like what \kcfa{} does with control-flow graphs.

Melski and Reps \cite{journal/tcs/00/melski/cflReachab} reduced Heintze's
set-constraints \cite{diss/cmu/92/heintze} to an instance of 
context-free-language (\abbrev\ CFL) reachability, which they solve using
summarization.
Therefore, their solution has the same precision as \cfa{0}.

CFL reachability has also been used for points-to analysis of imperative 
higher-order languages.
For instance, Sridharan and Bod\'ik's points-to analysis for Java
\cite{conf/pldi/06/sridharan/pointsTo} uses CFL reachability to match writes
and reads to object fields.
Precise call/return matching is achieved only for programs without recursive
methods. 
Hind's survey \cite{conf/paste/01/hind/survey} discusses many other
variants of points-to analysis.

Debray and Proebsting \cite{journal/toplas/97/debray/tailcall} used ideas from
parsing theory to design an interprocedural analysis for first-order programs 
with tail calls.
They describe control-flow with a context-free grammar.
Then, the {\small FOLLOW} set of a procedure represents its possible 
return points.
Our approach is quite different on the surface, but similar in spirit; 
we handle tail calls by computing summaries transitively.

Mossin \cite{diss/diku/96/mossin} created a type-based flow analysis for 
functional languages, which uses poly\-mor\-phic subtyping for polyvariance.
The input to the analysis is a program $p$ in the simply-typed \lam-calculus 
with recursion.
First, the analysis annotates the types in $p$ with labels.
Then, it computes flow information by assigning labeled types to each expression
in $p$.
Thus, flow analysis is reduced to a type-inference problem.
The annotated type system uses let-polymorphism.
As a result, it can distinguish flows to different references of let- and 
letrec-bound variables.
In the following program, it finds that \tw{n2} is a constant.
\begin{center}
  \begin{minipage}{0.35\columnwidth}
    \begin{alltt}
(let* ((id (\lam(x) x))
       (n1 (id 1))
       (n2 (id 2)))
  (+ n1 n2)) \end{alltt}
  \end{minipage}
\end{center}
However, the type system merges flows to different references of \lam-bound
variables.
For instance, it cannot find that \tw{n2} is a constant in the \tw{app} example
of section \ref{sec:call-ret-mismatch}.
Mossin's algorithm runs in time $O(n^8)$.

Rehof and F\"ahndrich \cite{conf/popl/01/rehof/typeflow, 
journal/mscs/08/fahndrich/typeflow} used CFL reachability in an analysis that
runs in cubic time and has the same precision as Mossin's.
They also extended the analysis to handle polymorphism in the target language.
Around the same time, Gustavsson and Svenningsson 
\cite{conf/pado/01/gustavsson/constr} formulated a cubic version of Mossin's
analysis without using CFL reachability.
Their work does not deal with polymorphism in the target language.

Midtgaard and Jensen \cite{conf/icfp/09/midtgaard/directCallRet} created 
a flow analysis for direct-style higher-order programs that keeps track of
``return flow''.
They point out that continuations make return-point information explicit
in \cps{} and show how to recover this information in direct-style programs.
Their work does not address the issue of unbounded call/return matching.

Earl \etal{} followed up on \cfat{} with a pushdown analysis that does not use 
frames \cite{conf/sfp/10/earl/pdcfa}.
Rather, it allocates all bindings in the heap with context, in the style of 
$k$-CFA \cite{diss/cmu/91/olin}.
For $k = 0$, their analysis runs in time $O(n^6)$, where $n$ is the size of the
program.
Like all pushdown-reachability algorithms, Earl \etal's analysis records pairs
of states $(\cstato, \cstatw)$ where \cstatw{} is same-context reachable from
\cstato.
However, their algorithm does not classify states as entries, exits, calls, 
\etc.
This has two drawbacks compared to the tabulation algorithm.
First, they do not distinguish between path and summary edges.
Thus, they have to search the whole set of edges when they look for return 
points, even though only summaries can contribute to the search.
More importantly, path edges are only a small subset of the set $S$ of all edges
between same-context reachable states.
By not classifying states, their algorithm maintains the whole set $S$, not just
the path edges.
In other words, it records edges whose source is not an entry.
In the graph of \tw{len}, some of these edges are \tuple{6,8}, \tuple{6,13}, 
\tuple{7,11}.
Such edges slow down the analysis and do not contribute to call/return matching,
because they cannot evolve into summary edges.

In CFA2, it is possible to disable the use of frames by classifying each 
reference as a heap reference.
The resulting analysis has similar precision to Earl \etal's analysis for $k=0$.
We conjecture that this variant is not a viable alternative in practice, because
of the significant loss in precision.

Might and Shivers \cite{conf/icfp/06/might/gcfa} proposed 
\cfa{\Gamma} (abstract garbage collection) and \cfa{\mu} (abstract counting) 
to increase the precision of \kcfa.
\cfa{\Gamma} removes unreachable bindings from the variable environment,
and \cfa{\mu} counts how many times a variable is bound during the analysis.
The two techniques combined reduce the number of spurious flows and give precise
environment information.
Stack references in \cfat{} have a similar effect, because different calls to 
the same function use different frames.
However, we can utilize \cfa{\Gamma} and \cfa{\mu} to improve precision in the
heap.

Recently, Kobayashi \cite{conf/popl/09/kobayashi/hors} proposed a way to 
statically verify properties of typed higher-order programs using 
model-checking.
He models a program by a higher-order recursion scheme $\mathcal{G}$,
expresses the property of interest in the modal $\mu$-calculus
and checks if the infinite tree generated by $\mathcal{G}$ satisfies the 
property.
This technique can do flow analysis, since flow analysis can be encoded as a 
model-checking problem.
The target language of this work is the simply-typed lambda calculus.
Programs in a Turing-complete language must be approximated in the simply-typed
lambda calculus in order to be model-checked.

\section{Conclusions\label{sec:conclusions}}

\noindent In this paper we propose \cfat{}, a pushdown model of higher-order programs,
and prove it correct.
\cfat{} provides precise call/return matching 
and has a better approach to variable binding than \kcfa{}.
Our evaluation shows that \cfat{} gives more accurate data-flow
information than \cfa{0} and \cfa{1}.

Stack lookups make \cfat{} polyvariant because different calls to the same 
function are analyzed in different environments.
We did not add polyvariance in the heap to keep the presentation simple.
Heap polyvariance is \emph{orthogonal} to call/return matching; integrating
existing techniques \cite{diss/cmu/91/olin, conf/ecoop/95/agesen/cpa, 
journal/toplas/98/wright/polysplit} in \cfat{} should raise no difficulties.
For example, \cfat{} can be extended with call-strings polyvariance, like \kcfa,
to produce a family of analyses $\cfat.0$, $\cfat.1$ and so on.
Then, any instance of $\cfat.k$ would be strictly more precise than the 
corresponding instance of \kcfa.

We believe that pushdown models are a better tool for higher-order flow analysis
than control-flow graphs,
and are working on providing more empirical support to this thesis.
We plan to use \cfat{} for environment analysis and stack-related optimizations.
We also plan to add support for \tw{call/cc} in \cfat{}. 

\paragraph{\myp{Acknowledgements}}
We would like to thank Danny Dub\'e for discovering the stack-filtering exploit
and for giving us permission to include it here.
Thanks also to Mitch Wand and the anonymous reviewers for their helpful comments
on the paper.

\bibliographystyle{plain}
\bibliography{dimvar-refs}
\clearpage
\appendix

\section{}

\noindent We use the notation \proj{i}{\tuple{e_1,\dots,e_n}} to retrieve the $i^{\mr{th}}$
element of a tuple \tuple{e_1, \dots, e_n}.
Also, we write $\getlab{g}$ to get the label of a term $g$.

In section \ref{sec:basics}, we mentioned that labels in a program can be split
into disjoint sets according to the innermost user lambda that contains them.
The ``label to label'' map \ltol{\psi} returns the labels that are in the same 
set as $\psi$.
For example, in the program \tw{(\ilam{1}(x k1) (k1 (\ilam{2}(y k2) (x y %
(\ilam{3}(u) (x u k2)\(\sp{4}\)))\(\sp{5}\)))\(\sp{6}\))}, these sets are 
\mset{1, 6} and \mset{2, 3, 4, 5}, so we know $\ltol{4} = \mset{2, 3, 4, 5}$ 
and $\ltol{6} = \mset{1, 6}$.

\begin{defi}
For every term $g$, the map \bv{g} returns the variables bound by lambdas 
which are subterms of $g$.
The map has a simple inductive definition: \\
$\bv{\denot{\anylam}} =
\mset{v_1, \dots, v_n} \cup \bv{\mcall}$ \\
$\bv{\denot{\anycall}} = \bv{g_1} \cup \dots \cup \bv{g_n}$ \\
$\bv{v} = \emptyset$
\qed
\end{defi}

We assume that \cfat{} works on an alphatized program, \ie, a program where all
variables have distinct names.
Thus, if \anylam{} is a term in such a program, we know that no other lambda in
that program binds variables with names $v_1, \dots, v_n$.
(During execution of \cfat, we do not rename any variables.)
The following lemma is a simple consequence of alphatization.
\begin{lem}\label{lem:envs-have-no-junk}
A concrete state \cstat{} has the form $(\dots, \cvenv, t)$.
\begin{enumerate}[\em(1)]
\item
  For any closure $(\mlam, \cbenv) \in \range{\cvenv}$, it holds that 
  $\dom{\cbenv} \cap \bv{\mlam} = \emptyset$.
\item
  If \cstat{} is an \deval{} with call site \mcall{} and environment 
  \cbenv, then $\dom{\cbenv} \cap \bv{\mcall} = \emptyset$.
\item
  If \cstat{} is an \dapply, for any closure $(\mlam, \cbenv)$ in operator or 
  argument position, then $\dom{\cbenv} \cap \bv{\mlam} = \emptyset$.
\end{enumerate}
\end{lem}
\proof
We show that the lemma holds for the initial state \initstate.
Then, for each transition $\cstat \cstep \cstat'$, we assume that \cstat{} 
satisfies the lemma and show that $\cstat'$ also satisfies it.
\begin{enumerate}[$\bullet$]
\item 
  \initstate{} is a \duapply{} of the form
  $((\mprog,\emptyset), (\mlam, \emptyset), \haltcont, \emptyset, \tuple{})$.
  Since \cvenv{} is empty, (1) trivially holds.
  Also, both closures have an empty environment so (3) holds.
\item
  The \labr{UEA} transition is: \\
  $(\denot{\ucall}, \cbenv, \cvenv, t) \cstep 
  (\mcproc, \cuarg, \ccarg, \cvenv, l::t)$ \\
  $\mcproc = \cbiga{f, \cbenv, \cvenv}$ \\
  $\cuarg = \cbiga{e, \cbenv, \cvenv}$ \\
  $\ccarg = \cbiga{q, \cbenv, \cvenv}$ 
  \\ \\
  The \cvenv{} doesn't change in the transition, so (1) holds for $\cstat'$.\\
  The operator is a closure of the form $(\mlam, \cbenv')$.
  We must show that $\dom{\cbenv'} \cap \bv{\mlam} = \emptyset$.
  If \islam{f}, then $\mlam = f$ and $\cbenv' = \cbenv$.
  Also, we know \\
  $\dom{\cbenv} \cap \bv{\denot{\ucall}} = \emptyset$ \\
  $\Rightarrow\; \dom{\cbenv} \cap (\bv{f}\cup\bv{e}\cup\bv{q}) = \emptyset$ \\
  $\Rightarrow\; \dom{\cbenv} \cap \bv{f} = \emptyset$. \\
  If \isvar{f}, then $(\mlam, \cbenv') \in \range{\cvenv}$, so we get the 
  desired result because \cvenv{} satisfies (1). \\
  Similarly for \cuarg{} and \ccarg.
\item
  The \labr{UAE} transition is: \\
  $(\mcproc, \cuarg, \ccarg, \cvenv, t) \cstep (\mcall, \cbenv', \cvenv', t)$ \\
  $\mcproc \equiv \tuple{\denot{\ulam}, \cbenv}$ \\
  $\cbenv' = \cbenv \onemap{u}{t} \onemap{k}{t}$ \\
  $\cvenv' = \cvenv \onemap{(u, t)}{\cuarg} \onemap{(k, t)}{\ccarg}$ 
  \\ \\
  To show (1) for $\cvenv'$, it suffices to show that \cuarg{} and \ccarg{}
  don't violate the property.
  The user argument \cuarg{} is of the form $(\mlam_1, \cbenv_1)$.
  Since \cstat{} satisfies (3), we know $\dom{\cbenv_1} \cap \bv{\mlam_1} =
  \emptyset$, which is the desired result.
  Similarly for \ccarg.

  Also, we must show that $\cstat'$ satisfies (2).
  We know $\mset{u, k} \cap \bv{\mcall} = \emptyset$ because the program is
  alphatized.
  Also, from property (3) for \cstat{} we know $\dom{\cbenv} \cap 
  \bv{\denot{\ulam}} = \emptyset$, which implies 
  $\dom{\cbenv} \cap \bv{\mcall} = \emptyset$.
  We must show \\
  $\dom{\cbenv'} \cap \bv{\mcall} = \emptyset$ \\
  $\Leftrightarrow\; 
  (\dom{\cbenv} \cup \mset{u, k}) \cap \bv{\mcall} = \emptyset$ \\
  $\Leftrightarrow\; (\dom{\cbenv} \cap \bv{\mcall}) \cup 
  (\mset{u, k} \cap \bv{\mcall}) = \emptyset$ \\
  $\Leftrightarrow\; \emptyset \cup \emptyset = \emptyset$.
\item
  Similarly for the other two transitions. \qed
\end{enumerate}

\begin{thm}[Simulation]
If $\cstat \cstep \cstat'$ and $\ctoa{\cstat} \sqsubseteq \astat$, then 
there exists $\astat'$ such that $\astat \astep \astat'$
and $\ctoa{\cstat'} \sqsubseteq \astat'$.
\end{thm}
\proof
By cases on the concrete transition. 
\begin{enumerate}[$\bullet$]
  \item[a)]
    Rule \labr{UEA} \\
    $(\denot{\ucall}, \cbenv, \cvenv, t) \cstep 
    (\mcproc, \cuarg, \ccarg, \cvenv, l::t)$ \\
    $\mcproc = \cbiga{f, \cbenv, \cvenv}$ \\
    $\cuarg = \cbiga{e, \cbenv, \cvenv}$ \\
    $\ccarg = \cbiga{q, \cbenv, \cvenv}$ \\ 
    \\
    Let $\tsenv = \tostack{\ltov{l}}{\cbenv}{\cvenv}$.
    Since $\ctoa{\cstat} \sqsubseteq \astat$, \astat{} is of the form 
    $(\denot{\ucall}, \stenv, \henv)$, where $\ctoa{\cvenv} \sqsubseteq \henv$
    and $\tsenv \sqsubseteq \stenv$. \\
    \\
    The abstract transition is \\
    $(\denot{\ucall}, \stenv, \henv) \astep 
    (f', \auarg, \acarg, \stenv', \henv)$ \\
    $f' \in \ubiga{f, l, \stenv, \henv}$ \\
    $\auarg = \ubiga{e, l, \stenv, \henv}$ \\
    $\acarg = \kbiga{q, \stenv}$ \\
    $\stenv' =
    \begin{cases}
      \pop{\stenv} & \isvar{q} \\
      \stenv & \islam{q} \land (\inheap{l, f} \lor \islam{f}) \\
      \stenv\onemap{f}{\mset{f'}} & \islam{q} \land \instack{l, f}
    \end{cases}$ \\
    \\
    State \astat{} has many possible successors, one for each lambda in 
    \ubiga{f, l, \stenv, \henv}.
    We must show that one of them is a state $\astat'$ such that 
    $\ctoa{\cstat'} \sqsubseteq \astat'$.

    The variable environment and the heap don't change in the transitions, so 
    for $\cstat'$ and $\astat'$ we know that $\ctoa{\cvenv} \sqsubseteq \henv$.
    We must show $\proj{1}{\mcproc} = f'$, $\ctoa{\cuarg} \sqsubseteq \auarg$,
    $\ctoa{\ccarg} \sqsubseteq \acarg$ and $\tsenv' \sqsubseteq \stenv'$,
    where $\tsenv'$ is the stack of \ctoa{\cstat'}. \\
    We first show $\proj{1}{\mcproc} = f'$, by cases on $f$:
    \begin{enumerate}[$\bullet$]
    \item 
      \islam{f} \\
      Then, $\mcproc = (f, \cbenv)$ and $f' \in \mset{f}$, so $f' = f$.
    \item
      \instack{l, f} \\
      Then, $\mcproc = \cvenv(f, \cbenv(f))$, a closure of the form 
      $(\mlam, \cbenv')$.
      Since $\tsenv(f) = \ctoa{\cvenv(f, \cbenv(f))}= \mset{\mlam}$ and
      $\tsenv \sqsubseteq \stenv$, we get $\mlam \in \stenv(f)$.
      So, we pick $f'$ to be \mlam.
    \item
      \inheap{l, f} \\
      Then, $\mcproc = \cvenv(f, \cbenv(f))$, a closure of the form 
      $(\mlam, \cbenv')$.
      Since $\ctoa{\cvenv} \sqsubseteq \henv$ and $\mlam \in \ctoa{\cvenv}(f)$,
      we get $\mlam \in \henv(f)$.
      So, we pick $f'$ to be \mlam.
    \end{enumerate}
    Showing $\ctoa{\cuarg} \sqsubseteq \auarg$ is similar. \\
    We now show $\ctoa{\ccarg} \sqsubseteq \acarg$, by cases on $q$:
    \begin{enumerate}[$\bullet$]
    \item 
      \islam{q} \\
      Then, $\ccarg=(q,\cbenv)$ and $\acarg=q$, so $\ctoa{c}\sqsubseteq\acarg$.
    \item 
      \isvar{q} and $\ccarg = \cvenv(q, \cbenv(q)) = \haltcont$ \\
      Then, $\tsenv(q) = \haltcont$.
      Since $\tsenv \sqsubseteq \stenv$, we get $\stenv(q) = \haltcont$.
      Thus, $\acarg = \haltcont$.
    \item
      \isvar{q} and $\ccarg = \cvenv(q, \cbenv(q)) = (\mlam, \cbenv')$ \\
      Similar to the previous case.
    \end{enumerate}
    It remains to show that $\tsenv' \sqsubseteq \stenv'$.
    We proceed by cases on $q$ and $f$: 
    \begin{enumerate}[$\bullet$]
    \item 
      \isvar{q} and $\ccarg = \cvenv(q, \cbenv(q)) = \haltcont$ \\
      Then, $\tsenv' = \tuple{}$.
      By $\tsenv \sqsubseteq \stenv$, we know that \tsenv{} and \stenv{} have
      the same size.
      Also, $\stenv' = \pop{\stenv}$, thus $\stenv' = \tuple{}$.
      Therefore, $\tsenv' \sqsubseteq \stenv'$.
    \item
      \isvar{q} and $\ccarg = \cvenv(q, \cbenv(q)) = (\mlam, \cbenv')$ \\
      By Fig.~\ref{fig:conc-to-abs}, we know that $\tsenv' = 
      \tostack{\ltov{\getlab{\mlam}}}{\cbenv'}{\cvenv} = \pop{\tsenv}$.
      Also, $\stenv' = \pop{\stenv}$.
      Thus, to show $\tsenv' \sqsubseteq \stenv'$ it suffices to show
      $\pop{\tsenv} \sqsubseteq \pop{\stenv}$, which holds because
      $\tsenv \sqsubseteq \stenv$.
    \item
      $\islam{q} \land (\islam{f} \lor \inheap{l, f})$ \\
      Then, $\tsenv' = \tsenv$ and $\stenv' = \stenv$, so $\tsenv' \sqsubseteq 
      \stenv'$.
    \item
      $\islam{q} \land \instack{l, f}$ \\
      By $\ltov{\getlab{q}} = \ltov{l}$, we get that $\tsenv' = \tsenv$.
      Also, $\mcproc = \cvenv(f, \cbenv(f))$, a closure of the form 
      $(\mlam, \cbenv')$.
      We pick $f'$ to be \mlam.
      The stack of $\astat'$ is $\stenv' = \stenv\onemap{f}{\mset{\mlam}}$.
      Since $\pop{\tsenv} \sqsubseteq \pop{\stenv}$, we only need to show that
      the top frames of $\tsenv'$ and $\stenv'$ are in $\sqsubseteq$.
      For this, it suffices to show that $\tsenv'(f) \sqsubseteq \stenv'(f)$
      which holds because $\tsenv'(f) = \tsenv(f) = \mset{\mlam}$.
    \end{enumerate}
  $\phantom{want some space between cases a) and b) }$
  \item[b)]
    Rule \labr{UAE} \\
    $(\mcproc, \cuarg, \ccarg, \cvenv, t) \cstep (\mcall, \cbenv', \cvenv', t)$
    \\
    $\mcproc \equiv \tuple{\denot{\ulam}, \cbenv}$\\
    $\cbenv' = \cbenv \onemap{u}{t} \onemap{k}{t}$ \\
    $\cvenv' = \cvenv \onemap{(u, t)}{\cuarg} \onemap{(k, t)}{\ccarg}$  \\
    \\
    Let $\tsenv = 
    \begin{cases}
      \tuple{} & \ccarg = \haltcont \\
      \tostack{\ltov{\getlab{\mlam}}}{\cbenv_1}{\cvenv} 
      & \ccarg = (\mlam, \cbenv_1)
    \end{cases}$ \\
    Since $\ctoa{\cstat} \sqsubseteq \astat$, \astat{} is of the form
    $(\denot{\ulam}, \auarg, \acarg, \stenv, \henv)$, 
    where $\ctoa{\cuarg} \sqsubseteq \auarg$, $\ctoa{\ccarg} = \acarg$,
    $\tsenv \sqsubseteq \stenv$ and $\ctoa{\cvenv} \sqsubseteq \henv$.
    \\ \\
    The abstract transition is \\
    $(\denot{\ulam}, \auarg, \acarg, \stenv, \henv) \astep 
    (\mcall, \stenv', \henv')$ \\
    $\stenv' = \push{\onemap{u}{\auarg} \onemap{k}{\acarg}} {\stenv}$ \\
    $\henv' = 
    \begin{cases}
      \henv\join\onemap{u}{\auarg} & \inheap{u} \\
      \henv & \instack{u}
    \end{cases}$ \\
    \\
    Let $\tsenv'$ be the stack of $\ctoa{\cstat'}$.
    The innermost user lambda that contains \mcall{} is \ilam{l}, therefore 
    $\tsenv' = \tostack{\ltov{l}}{\cbenv'}{\cvenv'}$.
    We must show that $\ctoa{\cstat'} \sqsubseteq \astat'$, \ie,
    $\tsenv' \sqsubseteq \stenv'$ and $\ctoa{\cvenv'} \sqsubseteq \henv'$.

    We assume that $\ccarg = (\mlam, \cbenv_1)$ and that \inheap{u} holds, the 
    other cases are simpler.
    In this case, $\ctoa{\cvenv'}$ is the same as $\ctoa{\cvenv}$ except that
    $\ctoa{\cvenv'}(u) = \ctoa{\cvenv}(u) \join \ctoa{\cuarg}$.
    Also, $\henv'(u) = \henv(u) \join \auarg$, thus $\ctoa{\cvenv'} 
    \sqsubseteq \henv'$.
    
    We know that $\cbenv'$ contains bindings for $u$ and $k$, and by lemma
    \ref{lem:envs-have-no-junk} it doesn't bind any variables in \bv{\mcall}.
    Since $\ltov{l} \setminus \mset{u, k} = \bv{\mcall}$, $\cbenv'$ doesn't bind
    any variables in $\ltov{l} \setminus \mset{u, k}$.
    Thus, the top frame of $\tsenv'$ is 
    $\onemap{u}{\ctoa{\cuarg}} \onemap{k}{\ctoa{\ccarg}}$.
    The top frame of $\stenv'$ is $\onemap{u}{\auarg} \onemap{k}{\acarg}$,
    therefore the frames are in $\sqsubseteq$.
    To complete the proof of $\tsenv' \sqsubseteq \stenv'$, we must show that
    $\pop{\tsenv'} \sqsubseteq \pop{\stenv'}$ \\
    $\Leftrightarrow\; \pop{\tsenv'} \sqsubseteq \stenv$ \\
    $\Leftarrow\; \pop{\tsenv'} = \tsenv$. \\
    We know $\pop{\tsenv'}= \tostack{\ltov{\getlab{\mlam}}}{\cbenv_1}{\cvenv'}$,
    $\tsenv = \tostack{\ltov{\getlab{\mlam}}}{\cbenv_1}{\cvenv}$.
    By the temporal consistency of states (\confer{} \cite{diss/07/might/dcfa} 
    definition 4.4.5), \pop{\tsenv'} won't contain the two bindings born at time
    $t$ because they are younger than all bindings in $\cbenv_1$.
    This implies that $\pop{\tsenv'} = \tsenv$.
    \\ $\phantom{want some space between the two cases}$
  \item[c)]
    Rule \labr{CEA} \\
    $(\denot{\qcall}, \cbenv, \cvenv, t) \cstep 
    (\mcproc, \cuarg, \cvenv, \gamma::t)$ \\
    $\mcproc = \cbiga{q, \cbenv, \cvenv}$ \\
    $\cuarg = \cbiga{e, \cbenv, \cvenv}$ 
    \\ \\
    Let $\tsenv = \tostack{\ltov{\gamma}}{\cbenv}{\cvenv}$.
    Since $\ctoa{\cstat} \sqsubseteq \astat$, \astat{} is of the form 
    $(\denot{\qcall}, \stenv, \henv)$, where $\ctoa{\cvenv} \sqsubseteq \henv$
    and $\tsenv \sqsubseteq \stenv$.
    The abstract transition is \\
    $(\denot{\qcall}, \stenv, \henv) \astep (q', \auarg, \stenv', \henv)$ 
    \rule{0cm}{0.6cm}\\
    $q' = \kbiga{q, \stenv}$ \\
    $\auarg = \ubiga{e, \gamma, \stenv, \henv}$ \\
    $\stenv' = 
    \begin{cases}
      \pop{\stenv} & \isvar{q} \\
      \stenv & \islam{q}
    \end{cases}$ \\
    \\ 
    Let $\tsenv'$ be the stack of \ctoa{\cstat'}.
    We must show that $\ctoa{\cstat'} \sqsubseteq \astat'$, \ie,
    $\ctoa{\mcproc} = q'$, $\ctoa{\cuarg} \sqsubseteq \auarg$, 
    and $\tsenv' \sqsubseteq \stenv'$. \\
    We first show $\ctoa{\mcproc} = q'$, by cases on $q$:
    \begin{enumerate}[$\bullet$]
    \item 
      \islam{q} \\
      Then, $\mcproc = (q, \cbenv)$ and $q' = q$.
      Thus, $\ctoa{\mcproc} = q'$.
    \item
      \isvar{q} and $\mcproc = \cvenv(q, \cbenv(q)) = (\mlam, \cbenv_1)$ \\
      Since $q \in \ltov{\gamma}$ we get $\tsenv(q) = \mlam$.
      From the latter and $\tsenv \sqsubseteq \stenv$, we get $\stenv(q) =
      \mlam$, which implies $q' = \mlam$, which implies $\ctoa{\mcproc} = q'$.
    \item
      \isvar{q} and $\mcproc = \cvenv(q, \cbenv(q)) = \haltcont$ \\
      Similar to the previous case.
    \end{enumerate}
    Showing $\ctoa{\cuarg} \sqsubseteq \auarg$ is similar, by cases on $e$. \\
    Last, we show $\tsenv' \sqsubseteq \stenv'$, by cases on $q$:
    \begin{enumerate}[$\bullet$]
    \item 
      \islam{q} \\
      Then, $\stenv' = \stenv$.
      Also, $\tsenv' = \tostack{\ltov{\getlab{q}}}{\cbenv}{\cvenv}$ and
      $\ltov{\getlab{q}} = \ltov{\gamma}$.
      Thus, $\tsenv' = \tsenv$, which implies $\tsenv' \sqsubseteq \stenv'$.
    \item
      \isvar{q} and $\mcproc = \cvenv(q, \cbenv(q)) = (\mlam, \cbenv_1)$ \\
      Then, $\tsenv' = \tostack{\ltov{\getlab{\mlam}}}{\cbenv_1}{\cvenv} = 
      \pop{\tsenv}$ and $\stenv' = \pop{\stenv}$.
      To show $\tsenv' \sqsubseteq \stenv'$, it suffices to show $\pop{\tsenv} 
      \sqsubseteq \pop{\stenv}$, which holds by $\tsenv \sqsubseteq \stenv$.
    \item
      \isvar{q} and $\mcproc = \cvenv(q, \cbenv(q)) = \haltcont$ \\
      Similar to the previous case.
    \end{enumerate}
    $\phantom{want some space between the two cases}$
  \item[d)]
    Rule \labr{CAE} \\ 
    This case requires arguments similar to the previous cases.
    \qed
\end{enumerate}

\begin{lem} \label{lem:stack-eval}
  On an \daeval{}-to-\daapply{} transition, the stack below the top frame is 
  irrelevant. 
  Formally,
  \begin{enumerate}[$\bullet$]
  \item 
    If $(\denot{\ncall}, \tfenv::\stenv, \henv) \astep
    (\mulam, \auarg, \mlam, \tfenv'::\stenv, \henv)$ then for any $\stenv'$, \\
    $(\denot{\ncall}, \tfenv::\stenv', \henv) \astep
    (\mulam, \auarg, \mlam, \tfenv'::\stenv', \henv)$
  \item
    If $(\denot{\tcall}, \tfenv::\stenv, \henv) \astep
    (\mulam, \auarg, \acarg, \stenv, \henv)$
    then for any $\stenv'$, \\
    $(\denot{\tcall}, \tfenv::\stenv', \henv) \astep
    (\mulam, \auarg, \acarg, \stenv', \henv)$
  \item
    Similarly for rule \labar{CEA}.
    \qed
  \end{enumerate}
\end{lem}

\begin{lem} \label{lem:stack-apply}
  On an \daapply{}-to-\daeval{} transition, the stack is irrelevant.
  Formally, 
  \begin{enumerate}[$\bullet$]
  \item 
    If $(\denot{\ulam}, \auarg, \acarg, \stenv, \henv) \astep
    (\mcall, \onemap{u}{\auarg}\onemap{k}{\acarg}::\stenv, \henv')$
    then for any $\stenv'$, \\
    $(\denot{\ulam}, \auarg, \acarg, \stenv', \henv) \astep 
    (\mcall, \onemap{u}{\auarg}\onemap{k}{\acarg}::\stenv', 
    \henv')$
  \item
    Similarly for rule \labar{CAE}, where $\stenv'$ is any non-empty stack.
    \qed
  \end{enumerate}
\end{lem}

\newcommand{\entry}{\ensuremath{\astat_e}}
\begin{defi}[Push Monotonicity]\label{def:push-monotonic} ~\\
  Let $p \equiv \entry \asteps \astat$ 
  where \entry{} is an entry with stack $\stenv_e$.
  The path $p$ is \tb{push monotonic} iff every transition $\astato \astep 
  \astatw$ satisfies the following property:
  \begin{quote}
    If the stack of \astato{} is $\stenv_e$ then the transition can only push
    the stack, it cannot pop or modify the top frame.
  \end{quote}
  \qed
\end{defi}

\noindent
Push monotonicity is a property of paths, not of individual transitions.
A push monotonic path can contain transitions that pop, as long as the stack 
never shrinks below the stack of the initial state of the path.
The following properties are simple consequences of push monotonicity.
\begin{property}\label{prop:stack-suffix}
The stack of the first state in a push-monotonic path is a suffix of the stack
of every other state in the path.
\end{property}
\begin{property}
In a push-monotonic path, the number of pushes is greater than or equal to the
number of pops.
\end{property}

\noindent
The following lemma associates entries with ``same-level reachable'' states.
A state \astat{} is same-level reachable from an entry \entry{} if it is
in the procedure whose entry is \entry{} or if it is in some procedure
that can be reached from \entry{} through tail calls, \ie, without growing the
stack.

\begin{lem}[Same-level reachability]\label{lem:same-level} ~ \\
  Let $\entry = (\denot{\ulam}, \auarg, \acarg, \stenv_e, \henv_e)$, 
  $\astat = (\dots, \stenv, \henv)$, 
  and $p \equiv \entry \asteps \astat$ where $\entry \in \corens{\astat}$. 
  Then,
  \begin{enumerate}[\em(1)]
  \item\label{lem:same-level,case:entry}
    If \astat{} is an entry, $\stenv = \stenv_e$.
  \item\label{lem:same-level,case:non-entry}
    If \astat{} is not an entry,
    \begin{enumerate}[\em(a)]
    \item\label{lem:same-level,case:non-entryo}
      $\stenv$ is of the form $\tfenv :: \stenv_e$, for some frame \tfenv.
    \item\label{lem:same-level,case:non-entryw}
      there exists $k'$ such that $\tfenv(k') = \acarg$.
    \item\label{lem:same-level,case:non-entryh}
      if $\entry = \coren{\astat}$ then 
      $\dom{\tfenv} \subseteq \ltov{l}$, $\tfenv(u) \sqsubseteq \auarg$ and
      $\tfenv(k) = \acarg$. \\
      Moreover, if \astat{} is an \daeval{} over call site $\psi$ 
      then $\psi \in \ltol{l}$,
      and if \astat{} is a \dacapply{} over \tw{(\ilam{\gamma}($u'$)$\mcall'$)} 
      then $\gamma \in \ltol{l}$.
    \end{enumerate}
  \item\label{lem:same-level,case:push-monot}
    $p$ is push monotonic.
  \end{enumerate}
\end{lem}
\proof
By induction on the length \sizeof{p} of $p$.
Note that (\ref{lem:same-level,case:push-monot}) follows from the form of 
the stack in (\ref{lem:same-level,case:entry}) 
and (\ref{lem:same-level,case:non-entry}),
so we won't prove it separately. \\
Basecase: \\
If $\sizeof{p} = 0$, then $\astat{} = \entry$ so $\stenv = \stenv_e$. \\
Inductive step: \\
If $\sizeof{p} > 0$, there are two cases; either 
$\entry = \coren{\astat}$ or $\entry \neq \coren{\astat}$.
\begin{enumerate}[$\bullet$]
\item[a)]
  $\entry = \coren{\astat}$ \\
  Since $\sizeof{p} > 0$, \astat{} is not an entry, so the second or the
  third branch of the definition of \dcoren{} determine the shape of $p$.
  \begin{enumerate}[$\bullet$]
  \item[a1)]
    $p \equiv \entry \asteps \astat' \astep \astat$ \\
    Here, the predecessor $\astat'$ of \astat{} is not a \daceval{} exit,
    and $\entry = \coren{\astat'}$.
    We proceed by cases on $\astat'$.
    Note that $\astat'$ cannot be a \daueval{} 
    because then \astat{} is an entry, so $\astat = \coren{\astat}$,
    and our assumption that $\entry = \coren{\astat}$ breaks.
    \begin{enumerate}[$\bullet$]
    \item[a1.1)]
      $\astat'$ is an inner \daceval \\
      Then, $\astat' =
      (\denot{\slp\slp \ilam{\gamma}\slp u'\srp\mcall'\srp \, e'\srp^{\gamma'}},
      \stenv', \henv')$.
      By \ih, $\stenv' = \tfenv'::\stenv_e$, \\
      $\dom{\tfenv'} \subseteq \ltov{l}$,
      $\tfenv'(u) \sqsubseteq \auarg$,
      $\tfenv'(k) = \acarg$ and
      $\gamma' \in \ltol{l}$.
      By the abstract semantics, $\astat = 
      (\denot{\slp \ilam{\gamma}\slp u'\srp\mcall'\srp}, \auarg', \stenv', 
      \henv')$
      where $\auarg' = \ubiga{e', \gamma', \stenv', \henv'}$.
      We know that $\gamma \in \ltol{l}$ 
      because $\gamma' \in \ltol{l}$.
      Also, the stack is unchanged in the transition.
      Thus, (\ref{lem:same-level,case:non-entryo}),
      (\ref{lem:same-level,case:non-entryw})
      and (\ref{lem:same-level,case:non-entryh})
      hold for \astat.
    \item[a1.2)]
      $\astat'$ is a \dacapply \\
      Then, $\astat' =
      (\denot{\slp \ilam{\gamma}\slp u'\srp\mcall'\srp}, 
      \auarg', \stenv', \henv')$.
      By \ih, $\stenv' = \tfenv'::\stenv_e$,
      $\dom{\tfenv'} \subseteq \ltov{l}$,
      $\tfenv'(u) \sqsubseteq \auarg$,
      $\tfenv'(k) = \acarg$ and
      $\gamma \in \ltol{l}$. \\
      By the abstract semantics, 
      $\astat = (\mcall', \stenv, \henv)$ where 
      $\stenv = \stenv'\onemap{u'}{\auarg'}$. \\
      So, $\stenv = \tfenv::\stenv_e$ which satisfies 
      (\ref{lem:same-level,case:non-entryo}).
      Also, $\tfenv = \tfenv'\onemap{u'}{\auarg'}$
      where $u' \in \ltov{l}$ because $\gamma \in \ltol{l}$,
      and $u' \neq u$ because the program is $\alpha$-tized.
      Thus,
      $\dom{\tfenv} = 
      \dom{\tfenv'} \cup \mset{u'} \subseteq \ltov{l}$,
      and $\tfenv(u) = \tfenv'(u) \sqsubseteq \auarg$,
      and $\tfenv(k) = \tfenv'(k) = \acarg$.
      Last, the label of $\mcall'$ is in \ltol{l} because 
      $\gamma \in \ltol{l}$.
    \item[a1.3)]
      $\astat'$ is a \dauapply \\
      Then, $\astat' = \entry$ because $\entry = \coren{\astat'}$.
      This case is simple.
    \end{enumerate}
  \item[a2)]
    \newcommand
        {\lamtwo}
        {\ensuremath{\slp\ilam{\gamma_2}\slp u_2\srp\mcall_2\srp}}
        \newcommand
            {\lamthree}
            {\ensuremath{\slp\ilam{l_3}\slp u_3 \, k_3\srp\mcall_3\srp}}
            $p \equiv \entry \astepso \astatw \astep \astath 
            \astepso \astat' \astep \astat$ \\
            Here, the third branch of the definition of \dcoren{} determines the
            shape of $p$,
            so $\astatw$ is a call, $\entry = \coren{\astatw}$,
            $\astat'$ is a \daceval{} exit and $\astath \in \corens{\astat'}$. \\
            By \ih{} for $\entry \astepso \astatw$ we get 
            $\astatw = (\denot{\slp f_2 \: e_2 \: \lamtwo\srp^{l_2}}, 
            \stenv_2, \henv_2)$,
            where $\stenv_2 \equiv \tfenv_2::\stenv_e$,
            $\dom{\tfenv_2} \subseteq \ltov{l}$,
            $\tfenv_2(u) \sqsubseteq \auarg$,
            $\tfenv_2(k) = \acarg$
            and $l_2 \in \ltol{l}$. \\
            By the abstract semantics for $\astatw \astep \astath$ we get \\
            $\astath = (\denot{\lamthree}, 
            \auarg_3, \acarg_3, \stenv_3, \henv_2)$, \\
            where $\denot{\lamthree} \in \ubiga{f_2, l_2, \stenv_2, \henv_2}$,\\
            $\auarg_3 = \ubiga{e_2, l_2, \stenv_2, \henv_2}$,
            $\acarg_3 = \denot{\lamtwo}$ and \\
            either $\stenv_3 = \stenv_2$, 
            if $\:(\islam{f_2} \lor \inheap{l_2, f_2})\:$ holds, \\
            or $\stenv_3 = \stenv_2\onemap{f_2}{\mset{\denot{\lamthree}}}$,
            if $\:\instack{l_2, f_2}\:$ holds.
            \begin{enumerate}[$\bullet$]
            \item[a2.1)]
              $\instack{l_2, f_2}$ \\
              Then,
              \mbox{$\stenv_3 \!=\! 
                \tfenv_2\onemap{f_2\!}{\!\mset{\denot{\lamthree}}} \!::\! \stenv_e$.} \\
              By \ih{} for $\astath \astepso \astat'$ we get 
              $\astat' = (\denot{\slp k' \: e'\srp^{\gamma'}}, 
              \stenv', \henv')$, \\
              where $\stenv' = \tfenv'::\stenv_3$ 
              and $\tfenv'(k') = \denot{\lamtwo}$. \\
              Thus, by the abstract semantics for $\astat' \astep \astat$
              we get \\
              $\astat = (\denot{\lamtwo}, \auarg', \stenv_3, \henv')$. \\
              Now, $\gamma_2 \in \ltol{l}$ follows from $l_2 \in \ltol{l}$. \\
              Also, $\stenv = \tfenv::\stenv_e$ 
              where $\tfenv = \tfenv_2\onemap{f_2}{\mset{\denot{\lamthree}}}$. \\
              Then, $\dom{\tfenv} = \dom{\tfenv_2} \cup \mset{f_2}
              \subseteq \ltov{l}$
              because \instack{l_2, f_2} implies $f_2 \in \ltov{l}$. \\
              Also, $\tfenv(k) = \tfenv_2(k) = \acarg$. 
              Last, we take cases depending on whether $u$ and $f_2$ are the
              same variable or not.
              \begin{enumerate}[$\bullet$]
              \item[$\bullet$]
                $u = f_2$ \\
                $\tfenv(u) = \mset{\denot{\lamthree}}$ 
                $\subseteq \ubiga{f_2, l_2, \stenv_2, \henv_2}$ 
                $= \stenv_2(f_2)$ 
                $= \tfenv_2(f_2)$
                $= \tfenv_2(u)$
                $ \sqsubseteq \auarg$
              \item[$\bullet$]
                $u \neq f_2$ \\
                $\tfenv(u) = \tfenv_2(u) \sqsubseteq \auarg$
              \end{enumerate}
            \item[a2.2)]
              $\islam{f_2} \lor \inheap{l_2, f_2}$ \\
              This case is simpler than the previous case because
              $\stenv_3 = \stenv_2$.
            \end{enumerate}
  \end{enumerate}
\item[b)]
  $\entry \neq \coren{\astat}$ (but $\entry \in \corens{\astat}$) \\
  Then, the second branch of the definition of \dcorens{} determines the
  shape of $p$; \\
  $p \equiv \entry \astepso \astato \astep \astatw \asteps \astat$,
  where \astato{} is a tail call, $\astatw = \coren{\astat}$
  and $\entry \in \corens{\astato}$. \\
  By \ih{} for $\entry \astepso \astato$ 
  we get $\astato = (\denot{\slp f_1 \, e_1 \, k_1\srp^{l_1}}, 
  \stenv_1, \henv_1)$, \\
  where $\stenv_1 = \tfenv_1::\stenv_e$, 
  $\tfenv_1(k_1) = \acarg$. \\
  By the abstract semantics, 
  $\astatw = (\denot{\slp \ilam{l_2}\slp u_2 \, k_2\srp \mcall_2\srp}, 
  \auarg_2, \acarg, \stenv_e, \henv_1)$.
  \begin{enumerate}[$\bullet$] 
  \item[b.1)]
    \astat{} is an entry \\
    Then, $\astat = \astatw$ because $\astatw = \coren{\astat}$.
    So, $\stenv = \stenv_e$.
  \item[b.2)]
    \astat{} is not an entry \\
    By \ih{} for $\astatw \asteps \astat$ we get
    $\stenv \equiv \tfenv::\stenv_e$ and 
    $\tfenv(k_2) = \acarg$.
    This is the desired result for $\entry \asteps \astat$.
    \qed
  \end{enumerate}
\end{enumerate}

\begin{lem}[Local simulation]\label{lem:localsim} ~\\
If $\astat \astep \astat'$ and $\succ{\atol{\astat}} \neq \emptyset$,
then $\atol{\astat'} \in \succ{\atol{\astat}}$.
\end{lem}
\proof
By cases on the abstract transition. \\
We only show the lemma for \labar{UEA}, the other cases are similar. \\
$(\denot{\ucall}, \stenv, \henv) \astep (f', \auarg, \acarg, \stenv', \henv)$ \\
$f' \in \ubiga{f, l, \stenv, \henv}$ \\
$\auarg = \ubiga{e, l, \stenv, \henv}$ \\
$\acarg = \kbiga{q, \stenv}$ \\
$\stenv' =
\begin{cases}
  \pop{\stenv} & \isvar{q} \\
  \stenv & \islam{q} \land (\inheap{l, f} \lor \islam{f}) \\
  \stenv\onemap{f}{\mset{f'}} & \islam{q} \land \instack{l, f}
\end{cases}$ \\ ~\\
A \daueval{} state has a successor only when its stack is not empty,
so $\stenv \equiv \tfenv::\stenv''$. \\
Thus, $\atol{\stenv} = 
\msetcomp{(v, \tfenv(v))}{v \in \dom{\tfenv} \,\land\, \mi{UVar}_?(v)}$. \\
Then, $\atol{\astat} = (\denot{\ucall}, \atol{\stenv}, \henv)$. 
Also, $\atol{\astat'} = (f', \auarg, \henv)$. \\
If suffices to show that $f' \in \aubiga{f, l, \atol{\stenv}, \henv}$ and 
$\auarg = \aubiga{e, l, \atol{\stenv}, \henv}$; but these hold because 
$\aubiga{v, \psi, \atol{\stenv}, \henv} = \ubiga{v, \psi, \stenv, \henv}$ 
is true for any $v$ ($\mi{uvar}$ or \mulam{}).
\qed

\begin{lem}[Converse of Local Simulation]\label{lem:localsim-converse} ~\\
  If $\lstat \lstep \lstat'$ then, 
  for any \astat{} such that $\lstat = \atol{\astat}$, 
  there exists a state $\astat'$ such that 
  $\astat \astep \astat'$ and $\lstat' = \atol{\astat'}$
  \qed
\end{lem}

\begin{lem}[Path decomposition]\label{lem:decomp} 
  Let $p \equiv \entry \asteps \astat$ be push monotonic and
  $\entry = (\denot{\ulam}, \auarg, \acarg, \stenv_e, \henv_e)$.
  \begin{enumerate}[$\bullet$]
  \item
    if \astat{} is a \dacapply{} of the form
    $(\acarg, \dots, \stenv_e, \dots)$
    then \coren{\astat} is not defined.
  \item
    Otherwise,
  \end{enumerate}
    \begin{enumerate}[\quad\em(1)]
    \item\label{lem:decomp,case:coren}
      \coren{\astat} is defined, \ie{},
      $p \equiv \entry \asteps \astato \asteps \astat$,
      where $\astato = \coren{\astat}$.
    \item\label{lem:decomp,case:corens}
      Regarding the set \corens{\astat}, $p$ can be in one of four forms
      \begin{enumerate}[\em(a)]
      \item\label{lem:decomp,case:corens1}
        $p \equiv \entry \asteps \astat$ 
        where $\entry = \coren{\astat}$ and $\corens{\astat} = \mset{\entry}$
      \item\label{lem:decomp,case:corens2}
        $p \equiv e_1 \astepso c_1 \astep \dots \astep e_k \astepso c_k \astep 
        \astato \asteps \astat$, $k > 0$,
        where $e_i$s are entries, $c_i$s are tail calls,
        $e_1 = \entry$, $e_i = \coren{c_i}$, $\astato = \coren{\astat}$
        and $\corens{\astat} = \mset{e_1, \dots, e_k, \astato}$
      \item\label{lem:decomp,case:corens3}
        $p \equiv \entry \astepso c \astep \astato \asteps \astat$ 
        where $c$ is a call, $\astato = \coren{\astat}$
        and $\corens{\astat} = \mset{\astato}$
      \item\label{lem:decomp,case:corens4}
        $p \equiv \entry \astepso c \astep 
        e_1 \astepso c_1 \astep \dots \astep e_k \astepso c_k 
        \astep \astato \asteps \astat$, $k > 0$,
        where $c$ is a call, $e_i$s are entries, $c_i$s are tail calls,
        $e_i = \coren{c_i}$, $\astato = \coren{\astat}$
        and $\corens{\astat} = \mset{e_1, \dots, e_k, \astato}$
      \end{enumerate}
    \end{enumerate}
\end{lem}
\proof
By induction on the length of $p$. \\
Basecase: $\entry \astep^0 \entry$ \\
Then, $\astat = \entry$
$\;\Rightarrow\; \entry = \coren{\astat}$
$\;\Rightarrow\; \corens{\astat} = \mset{\entry}$
$\;\Rightarrow\;$ (\ref{lem:decomp,case:corens1}) holds \\
\\
Inductive step: $\entry \asteps \astat' \astep \astat$ \\
Cases on $\astat'$:
\begin{enumerate}[$\bullet$]
\item[a)]
  $\astat'$ is a Call \\
  Then, \astat{} is an entry so $\coren{\astat} = \astat$. 
  Also, $\corens{\astat} = \mset{\astat}$
  so (\ref{lem:decomp,case:corens3}) holds.
\item[b)]
  $\astat'$ is a Tail Call \\
  Then, \astat{} is an entry so $\coren{\astat} = \astat$. \\
  To show (\ref{lem:decomp,case:corens}), we take cases on whether 
  (\ref{lem:decomp,case:corens1}), (\ref{lem:decomp,case:corens2}),
  (\ref{lem:decomp,case:corens3}) or (\ref{lem:decomp,case:corens4})
  holds for $\astat'$.
  \begin{enumerate}[b.1]
  \item[b.1)]
    (\ref{lem:decomp,case:corens1}) holds for $\astat'$, \ie, \\
    $p \equiv \entry \asteps \astat' \astep \astat$ where
    $\entry = \coren{\astat'}$ and $\corens{\astat'} = \mset{\entry}$.
    By the second branch of the definition of \dcorens{},
    $\corens{\astat'} \subseteq \corens{\astat}$.
    Hence, $\corens{\astat} = \mset{\entry, \astat}$,
    which implies that (\ref{lem:decomp,case:corens2}) holds for \astat.
  \item[b.2)]
    (\ref{lem:decomp,case:corens2}) holds for $\astat'$ \\
    By a similar argument, 
    we find that (\ref{lem:decomp,case:corens2}) holds for \astat.
  \item[b.3)]
    (\ref{lem:decomp,case:corens3}) holds for $\astat'$ \\
    By a similar argument, 
    we find that (\ref{lem:decomp,case:corens4}) holds for \astat.
  \item[b.4)]
    (\ref{lem:decomp,case:corens4}) holds for $\astat'$ \\
    By a similar argument, 
    we find that (\ref{lem:decomp,case:corens4}) holds for \astat.
  \end{enumerate}
\item[c)]
  $\astat'$ is a $\dacapply{} \equiv (\acarg, \dots, \stenv_e, \dots)$ \\
  Then, in the transition $\astat' \astep \astat$ 
  we modify the top frame of $\stenv_e$, 
  which means that $p$ isn't push monotonic.
  Thus, this case can't arise.
\item[d)]
  $\astat'$ is an inner \daceval{}
  or a $\dacapply \not\equiv (\acarg, \dots, \stenv_e, \dots)$ \\
  By \ih{},
  $p \equiv \ainitstate \asteps \astato \astepso \astat' 
  \astep \astat$,
  where $\astato = \coren{\astat'}$. \\
  By the second branch of the definition of \dcoren{}, 
  $\astato = \coren{\astat}$.\\
  To show (\ref{lem:decomp,case:corens}), we take cases on whether 
  (\ref{lem:decomp,case:corens1}),
  (\ref{lem:decomp,case:corens2}),
  (\ref{lem:decomp,case:corens3})
  or (\ref{lem:decomp,case:corens4})
  holds for $\astat'$.
  The reasoning is the same as in case (b).
\item[e)]
  $\astat'$ is a \daceval{} exit \\
  By \ih{},
  $p \equiv \entry \asteps \astato \astepso \astat' 
  \astep \astat$,
  where $\astato = \coren{\astat'}$. \\
  Cases on (\ref{lem:decomp,case:corens1}), (\ref{lem:decomp,case:corens2}),
  (\ref{lem:decomp,case:corens3}) or (\ref{lem:decomp,case:corens4})
  for $\astat'$.
  \begin{enumerate}[$\bullet$]
  \item[e.1)]
    (\ref{lem:decomp,case:corens1}) holds for $\astat'$, \ie \\
    $p \equiv \entry \astepso \astat' \astep \astat$
    where $\entry = \coren{\astat'}$. \\
    By lemma~\ref{lem:same-level}, 
    the stack of $\astat'$ is of the form $\tfenv::\stenv_e$ and
    $\tfenv(k) = \acarg$.
    Thus, $\astat' \equiv (\acarg, \dots, \stenv_e, \dots)$.
    The only way for \coren{\astat} to exist is by the third branch of the
    definition of \dcoren{}, since $\astat'$ is a \daceval{} exit.
    But there is no call leading to \entry,
    thus \coren{\astat} can't exist.

    Similarly when (\ref{lem:decomp,case:corens2}) holds for $\astat'$.
  \item[e.2)]
    (\ref{lem:decomp,case:corens3}) holds for $\astat'$, \ie \\
    $p \equiv \entry \astepso c \astep \astato \astepso \astat' \astep
    \astat$ where $c$ is a call and $\astato = \coren{\astat'}$. \\
    By \ih{}, \coren{c} exists so $p$ can be written 
    $p \equiv \entry \asteps \astatw \astepso c \astep \astato 
    \astepso \astat' \astep \astat$ where $\astatw = \coren{c}$.
    Then, by the third branch of the definition of \dcoren{}, 
    $\coren{\astat} = \coren{c} = \astatw$. 

    To show (\ref{lem:decomp,case:corens}) for \astat{} 
    we work as in the previous cases.
  \end{enumerate}
\item[f)]
  $\astat'$ is an Entry \\
  This case is simple.
  \qed
\end{enumerate}

\begin{lem}[Stack irrelevance]\label{lem:stack-irrel}  ~\\
  Let $p \equiv \astat_1 \astep \astat_2 \astep \dots \astep \astat_n$
  be push monotonic,
  where $\astat_1 = (\mulam, \auarg, \acarg, \stenv_e, \henv_e)$.
  Also, $\astat_n$ is not a \dacapply{} of the form 
  $(\acarg, \dots, \stenv_e, \dots)$.
  By property \ref{prop:stack-suffix}, the stack of each $\astat_i$ is of the
  form \app{\stenv_i, \stenv_e}.
  \\
  For an arbitrary stack $\stenv'$ and continuation $\acarg'$, 
  consider the sequence $p'$ of states 
  $\; \astat_1' \; \astat_2' \; \dots \; \astat_n' \;$ 
  where each $\astat_i'$ is produced by $\astat_i$ as follows:
  \begin{enumerate}[$\bullet$]
  \item 
    if $\astat_i$ is an entry with stack $\stenv_e$ then replace the 
    continuation argument with $\acarg'$ and the stack with $\stenv'$.
  \item
    if $\stenv_e$ is a proper suffix of the stack of $\astat_i$ then 
    the latter has the form \app{\stenv_i', \tuple{\frenv_i}, \stenv_e}
    for some stack $\stenv_i'$.
    Change $\stenv_e$ to $\stenv'$ and bind the continuation variable in 
    $\frenv_i$ to $\acarg'$.
  \end{enumerate}
  (Note: the map isn't total, but it should be defined for all states 
  in $p$.) \\
  Then, 
  \begin{enumerate}[$\bullet$]
  \item
    for any two states $\astat_i'$ and $\astat_{i+1}'$ in $p'$, it holds that
    $\astat_i' \astep \astat_{i+1}'$
  \item
    the path $p'$ is push monotonic
  \end{enumerate}
\end{lem}
\newcommand{\astatnmo}{\eh{\astat_{n-1}}}
\proof
By induction on the length of $p$. \\
The basecase is simple. \\
Inductive step:
$p = \astat_1 \asteps \astatnmo \astep \astat_n$ \\
By \ih, the transitions in the path $\astat_1' \asteps \astatnmo'$ 
are valid with respect to the abstract semantics and the path is push monotonic.
We must show that $(\astatnmo', \astat_n') \in \astep$ and that
$\astat_1' \asteps \astat_n'$ is push monotonic. \\
Cases on $\astatnmo$:
\begin{enumerate}[(1)]
\item \label{lem:stack-irrel,case:ueval}
  \astatnmo{} is a \daueval{}, of the form $(\denot{\ucall}, \stenv, \henv)$ \\
  By lemma \ref{lem:decomp}, \coren{\astatnmo} is defined and $p$ can be in 
  one of four forms.
  We consider only the first case, the rest are similar. \\
  Let $p \equiv \astat_1 \astepso \astatnmo \astep \astat_n$ 
  where $\astat_1 = \coren{\astatnmo}$. \\
  By lemma \ref{lem:same-level}, \stenv{} is of the form
  $\tfenv::\stenv_e$ and the continuation variable in \tfenv{} (call it $k$)
  is bound to \acarg.
  \begin{enumerate}[(a)]
  \item 
    $q$ is a variable \\
    By the abstract semantics we have that
    $\astat_n$ is $(\mulam_n, \auarg_n, \acarg, \stenv_e, h)$.
    Also, the state $\astatnmo'$ is 
    $(\denot{\ucall}, \tfenv\onemap{k}{\acarg'}::\stenv', \henv)$,
    and it transitions to
    $(\mulam_n, \auarg_n, \acarg', \stenv', h)$ which is $\astat_n'$.
  \item
    $q$ is a lambda and $f$ is a stack reference \\
    Then, $\astat_n$ is 
    $(\mulam_n, \auarg_n, q, \tfenv\onemap{f}{\mset{\mulam_n}}::\stenv_e, h)$.\\
    Also, the state $\astatnmo'$ is 
    $(\denot{\ucall}, \tfenv\onemap{k}{\acarg'}::\stenv', \henv)$,
    and it transitions to \\
    $(\mulam_n, \auarg_n, q, \tfenv\onemap{k}{\acarg'}\onemap{f}{\mset{\mulam_n}}::\stenv', h)$ 
    which is $\astat_n'$.
  \item
    $q$ is a lambda and $f$ is a heap reference \\
    Similarly.
  \end{enumerate}
\item
  \astatnmo{} is a \daceval{} exit \\
  By lemma \ref{lem:decomp}, \coren{\astatnmo} is defined and $p$ can be in 
  one of four forms.
  \begin{enumerate}
  \item 
    $p \equiv \astat_1 \astepso \astatnmo \astep \astat_n$ where
    $\astat_1 = \coren{\astatnmo}$ \\
    Then, by lemma \ref{lem:same-level} and the abstract semantics,
    it is easy to see that $\astat_n$ is of the form
    $(\acarg, \dots, \stenv_e, \dots)$.
    Thus, this case isn't possible.

    Similarly when $\astat_1 \neq \coren{\astatnmo}$ 
    but is in \corens{\astatnmo}.
  \item
    $p \equiv \astat_1 \astepso c \astep \entry' \astepso \astatnmo 
    \astep \astat$ where $\entry' = \coren{\astatnmo}$ and $c$ is a call: \\
    Then, \coren{c} is defined and its stack has $\stenv_e$ as a suffix.
    Hence, by lemma \ref{lem:same-level}, the stack of $c$ is bigger than
    $\stenv_e$ by at least a frame.
    Since the stack of $\entry'$ has the same size as the stack of $c$,
    the stack of \astatnmo{} is bigger than $\stenv_e$ by at least two frames.
    By lemma \ref{lem:stack-eval} we get the desired result.

    Similarly when $\entry' \neq \coren{\astatnmo}$ 
    but is in \corens{\astatnmo}.
  \end{enumerate}
\item
  \astatnmo{} is an inner \daceval{} \\
  Similarly to the previous cases.
\item
  \astatnmo{} is a \dauapply{} \\
  Lemma \ref{lem:decomp} gives the same four cases.
  We only consider one, the rest are similar. \\
  Let $p \equiv \astat_1 \astepso c \astep \astatnmo \astep \astat_n$ 
  where $c$ is a call. \\
  Then, \coren{c} is defined and its stack has $\stenv_e$ as a suffix.
  Hence, by lemma \ref{lem:same-level}, the stack of $c$ is bigger than
  $\stenv_e$ by at least a frame.  
  Since the stack of \astatnmo{} has the same size as the stack of $c$,
  we don't change the continuation argument in $\astatnmo'$.
  By lemma \ref{lem:stack-apply} we get the desired result.
\item
  \astatnmo{} is a \dacapply{} \\
  Similarly to the previous cases.
  \qed
\end{enumerate}
$\phantom{i}$\newline
\begin{thm}[Soundness] ~\\
If $p \equiv \ainitstate \asteps \astat$ then, after summarization:
\begin{enumerate}[$\bullet$]
  \item
    if $\astat$ is not a final state then
    $(\atol{\coren{\astat}}, \atol{\astat}) \in \seen$
  \item
    if $\astat$ is a final state then
    $\atol{\astat} \in \finals$
  \item
    if $\astat$ is a \daceval{} exit and $\astat' \in \corens{\astat}$
    then $(\atol{\astat'}, \atol{\astat}) \in \seen$
\end{enumerate}
\end{thm}
\proof
By induction on the length of $p$. \\
Basecase: $\ainitstate \astep^0 \ainitstate$ \\
Then, $(\ainitstate, \ainitstate) \in \seen$. \\ 
\\
Inductive step: $\ainitstate \asteps \astat' \astep \astat$ \\
Cases on $\astat$:
\begin{enumerate}[$\bullet$]
\item[a)]
  \astat{} is an Entry \\
  Then, $\coren{\astat} = \astat$.
  Also, $\astat'$ is a call or a tail call. \\
  By lemma~\ref{lem:decomp}, $p \equiv \ainitstate \asteps \astato 
  \astepso \astat' \astep \astat$, 
  where $\astato = \coren{\astat'}$. \\
  By \ih, $(\atol{\astato}, \atol{\astat'}) \in \seen$ which means that it 
  has been entered in \work{} and examined.
  By lemma~\ref{lem:localsim}, 
  $\atol{\astat} \in \succ{\atol{\astat'}}$
  so in line 10 or 22 $(\atol{\astat}, \atol{\astat})$ will be propagated.
\item[b)]
  \newcommand{\qcallp}{\ensuremath{\tw{(}q \, e\tw{)}^{\gamma'}}}
  \newcommand{\ulamone}{\ensuremath{\tw{(}\lambda_{l_1}\tw{(}u_1\,k_1\tw{)}\,\mcall_1\tw{)}}}
  \newcommand{\ucalltwo}{\ensuremath{\tw{(}f_2 \: e_2 \: q_2\tw{)}^{l_2}}}
  \astat{} is a \dacapply{} but not a final state \\
  Then, $\astat = (\denot{\clam}, \auarg, \stenv, \henv)$ 
  and $\astat' = (\denot{\qcallp}, \stenv', \henv)$.
  \begin{enumerate}[$\bullet$]
  \item[b.1)]
    \islam{q}, \ie{} $\astat'$ is an inner \daceval \\
    This case is simple.
  \item[b.2)]
    \isvar{q}, \ie{} $\astat'$ is a \daceval{} exit \\
    The path $\ainitstate \asteps \astat'\,$ satisfies part
    \ref{lem:decomp,case:corens} of lemma~\ref{lem:decomp}.
    It can't satisfy cases
    \ref{lem:decomp,case:corens1} or \ref{lem:decomp,case:corens2}
    because \astat{} would be a final state by lemma~\ref{lem:same-level}.
    Thus, it satisfies 
    \ref{lem:decomp,case:corens3} or \ref{lem:decomp,case:corens4}.
    Then, the path is of the form 
    $p \equiv \ainitstate \asteps \astato \astepso \astatw
    \astep \astath \astepso \astat' \astep \astat$ \\
    where \astatw{} is a call, $\astato = \coren{\astatw}$
    and $\astath \in \corens{\astat'}$. 
    Note that by the third branch of the definition of \dcoren{},
    $\astato = \coren{\astat}$.
    We must show that $(\atol{\astato}, \atol{\astat}) \in \seen$. 
    \\ \\
    The state \astato{} is an entry of the form 
    $\astato = (\denot{\ulamone}, \auarg_1, \acarg_1, 
    \stenv_1, \henv_1)$ \\
    The state \astatw{} is a call of the form 
    $\astatw = (\denot{\ucalltwo}, \stenv_2, \henv_2)$,
    where $q_2$ is a \mclam{}. \\
    Lemma~\ref{lem:same-level} for $\astato \astepso \astatw$ gives
    $\stenv_2 \equiv \tfenv_2::\stenv_1$.\\
    By the abstract semantics for $\astatw \astep \astath$, we get:\\
    $\astath = (\mulam, \auarg_3, q_2, \stenv_3, \henv_2)$, where 
    \\
    either $\stenv_3=\stenv_2$, if $(\islam{f_2}\lor\inheap{l_2, f_2})$ holds,\\
    or $\stenv_3 = \stenv_2\onemap{f_2}{\mset{\mulam}}$,
    if $\instack{l_2, f_2}$ holds. \\
    \ie{} $\stenv_3 = \tfenv_3::\stenv_1$, and \\
    $\tfenv_3 = 
    \begin{cases}
      \tfenv_2 & \islam{f_2} \lor \inheap{l_2, f_2} \\
      \tfenv_2\onemap{f_2}{\mset{\mulam}} & \instack{l_2, f_2}
    \end{cases}$ \\
    By lemma~\ref{lem:same-level} for $\astath \astepso \astat'$, 
    we get $\stenv' = \tfenv'::\stenv_3$ and $\tfenv'(q) = q_2$. \\
    Then, by the abstract semantics for $\astat' \astep \astat$, \\
    $q_2 = \denot{\clam}$, $\stenv = \stenv_3$, and 
    $\auarg = \ubiga{e, \gamma', \stenv', \henv}$. \\
    The above information will become useful when dealing with the local
    counterparts of the aforementioned states. 
    \\
    By \ih{}, $(\atol{\astath}, \atol{\astat'})$ was entered in \work{}
    (at line 25) and later examined at line 13.
    Note that $\astath \neq \ainitstate$ because \astatw{} is between them,
    therefore \tw{Final} will not be called at line 15. 
    \\
    Also by \ih{}, $(\atol{\astato}, \atol{\astatw})$ was entered in \work{}
    and later examined.
    Lemma~\ref{lem:localsim} implies that 
    $\atol{\astath} \in \succ{\atol{\astatw}}$ so
    $(\atol{\astato}, \atol{\astatw}, \atol{\astath})$ will go in \callers.
    We take cases on whether $(\atol{\astath}, \atol{\astat'})$ or
    $(\atol{\astato}, \atol{\astatw})$ was examined first by the algorithm.
    \begin{enumerate}[$\bullet$]
    \item[b.2.1)]
      $(\atol{\astato}, \atol{\astatw})$ was examined first \\
      Then, when $(\atol{\astath}, \atol{\astat'})$ is examined, 
      $(\atol{\astato}, \atol{\astatw}, \atol{\astath})$ is in \callers. \\
      Therefore, at line 18 we call 
      \tw{Update(}$\atol{\astato}, \atol{\astatw}, \atol{\astath}, 
      \atol{\astat'}$\tw{)}. \\
      By applying \atol{\cdot} to the abstract states we get \\
      $\atol{\astato} = (\denot{\ulamone}, \auarg_1, \henv_1)$ \\
      $\atol{\astatw} = (\denot{\ucalltwo}, \tfenv_2, \henv_2)$, \\
      where $q_2 = \denot{\clam}$. \\
      $\atol{\astath} = (\mulam, \auarg_3, \henv_2)$ \\
      $\atol{\astat'} = (\denot{\qcallp}, \tfenv', \henv)$, \\
      where $\tfenv'(q) = \denot{\clam}$. \\
      By looking at \tw{Update}'s code, we see that the return value is 
      $\aubiga{e, \gamma', \tfenv', \henv} = \ubiga{e, \gamma', \stenv', \henv}
      = \auarg$.
      The frame of the return state is \\
      $\begin{cases}
        \tfenv_2 & \islam{f_2} \lor \inheap{l_2, f_2} \\
        \tfenv_2\onemap{f_2}{\mset{\mulam}} & \instack{l_2, f_2}
      \end{cases}$ \\
      which is equal to $\tfenv_3$.
      The heap at the return state is \henv.
      Last, the continuation we are returning to is \denot{\clam}.
      Thus, the return state \lstat{} is equal to \atol{\astat},
      and we call \tw{Propagate(}\atol{\astato}, \atol{\astat}\tw{)},
      so $(\atol{\astato}, \atol{\astat})$ will go in \seen.
    \item[b.2.2)]
      $(\atol{\astath}, \atol{\astat'})$ was examined first \\
      Then, when $(\atol{\astato}, \atol{\astatw})$ is examined,
      $(\atol{\astath}, \atol{\astat'})$ is in \summary{},
      and at line 12 we call 
      \tw{Update(}$\atol{\astato}, \atol{\astatw}, \atol{\astath}, \atol{\astat'}$\tw{)}. \\
      Proceed as above.
    \end{enumerate}
  \end{enumerate}
\item[c)]
  \astat{} is a final state \\
  Then, $\astat = (\haltcont, \auarg, \tuple{}, \henv)$. 
  We must show that \atol{\astat} will be in \finals{} after the execution of
  the summarization algorithm. 
  By the abstract semantics for $\astat' \astep \astat$, 
  $\astat' = (\denot{\kcall}, \stenv', \henv)$, 
  where $\stenv' = \tfenv'::\tuple{}$, 
  $\tfenv'(k) = \haltcont$,
  and $\auarg = \ubiga{e, \gamma, \stenv', \henv}$. \\
  By \ih{} for $\ainitstate \asteps \astat'$, we know that 
  $(\atol{\ainitstate}, \atol{\astat'})$ was entered in \work{} and \summary{}
  sometime during the algorithm.
  When it was examined, the test at line 14 was true so we called
  \tw{Final(}$\atol{\astat'}$\srp.
  Hence, we insert $\lstat = (\haltcont, \aubiga{e, \gamma, \tfenv', \henv},
  \emptyset, \henv)$ in \finals.
  But, $\aubiga{e, \gamma, \tfenv', \henv} = \ubiga{e, \gamma, \stenv', \henv} 
  = \auarg$, hence $\lstat = \atol{\astat}$.
\item[d)]
  \astat{} is a \daceval{} exit \\
  By lemma~\ref{lem:decomp} for $\ainitstate \asteps \astat'$, 
  $p \equiv \ainitstate \asteps \astato \asteps \astat' \astep 
  \astat$,
  where $\astato = \coren{\astat'}$. 
  But $\astat'$ is not a \daceval{} exit (it is an \daapply{} state), 
  so by the second branch of the definition of \dcoren{} 
  we get $\astato = \coren{\astat}$. \\
  By \ih{}, $(\atol{\astato}, \atol{\astat'})$ is entered in \seen{} and
  \work{}; and examined at line 6.
  By lemma~\ref{lem:localsim}, $\atol{\astat} \in \succ{\atol{\astat'}}$ 
  so $(\atol{\astato}, \atol{\astat})$ will be propagated (line 7)
  and entered in \seen{} (line 25). \\
  We need to show that for every $\astat'' \in \corens{\astat}$,
  $(\atol{\astat''}, \atol{\astat})$ will be inserted in \seen{}.
  The path $\ainitstate \asteps \astat'$ satisfies
  part \ref{lem:decomp,case:corens} of lemma \ref{lem:decomp};
  proceed by cases:
  \begin{enumerate}[$\bullet$]
  \item[d.1)]
    $\ainitstate \asteps \astat'$ satisfies \ref{lem:decomp,case:corens1}\\
    Then, $\astato = \ainitstate$ and 
    $p \equiv \astato \asteps \astat' \astep \astat$
    and $\corens{\astat} = \mset{\astato}$.
    But we 've shown that $(\atol{\astato}, \atol{\astat})$ is entered in \seen.
  \item[d.2)]
    $\ainitstate \asteps \astat'$ satisfies \ref{lem:decomp,case:corens2}\\
    Then,
    $p \equiv e_1 \astepso c_1 \astep \dots \astep e_k \astepso c_k 
    \astep \astato \asteps \astat' \astep \astat$, 
    where $e_1 = \ainitstate$, $e_i$s are entries, $c_i$s are tail calls,
    $e_i = \coren{c_i}$, 
    $\corens{\astat'} = \mset{e_1, \dots, e_k, \astato}$.

    Hence, $\corens{\astat} = \mset{e_1, \dots, e_k, \astato}$.
    To show that $(\atol{e_k}, \atol{\astat})$ is entered in \seen,
    we proceed by cases on whether 
    $(\atol{e_k}, \atol{c_k})$ or 
    $(\atol{\astato}, \atol{\astat})$ was examined first by the algorithm.
    \begin{enumerate}[$\bullet$]
    \item[d.2.1)]
      $(\atol{e_k}, \atol{c_k})$ was examined first \\
      By lemma~\ref{lem:localsim},
      $\atol{\astato}$ is in $\succ{\atol{c_k}}$, hence
      $(\atol{e_k}, \atol{c_k}, \atol{\astato})$
      will go in \tcallers.
      Then, when $(\atol{\astato}, \atol{\astat})$ is examined, 
      in line 19 we will call \tw{Propagate}\slp 
      $\atol{e_k}, \atol{\astat}$\srp, so
      $(\atol{e_k}, \atol{\astat})$ will go in \seen.
    \item[d.2.2)]
      $(\atol{\astato}, \atol{\astat})$ was examined first \\
      When $(\atol{e_k}, \atol{c_k})$ is
      examined, $(\atol{\astato}, \atol{\astat})$ will be in \summary,
      and by lemma~\ref{lem:localsim} we know 
      $\atol{\astato} \in \succ{\atol{c_k}}$.
      Thus, in line 24 we will call \tw{Propagate} which will insert
      $(\atol{e_k}, \atol{\astat})$ in \seen.
    \end{enumerate}
    By repeating this process $k - 1$ times, we can show that all edges
    $(\atol{e_i}, \atol{\astat})$ go in \seen.
  \item[d.3)]
    $\ainitstate \asteps \astat'$ satisfies \ref{lem:decomp,case:corens3}
    or \ref{lem:decomp,case:corens4} \\
    These cases are similar to the previous cases.
    The only difference is that now \ainitstate{} is not in
    \corens{\astat'} (which doesn't change the proof).
  \end{enumerate}
\item[e)]
  \astat{} is a Tail Call (thus an exit) \\
  By lemma~\ref{lem:decomp} for $\ainitstate \asteps \astat'$, 
  $p \equiv \ainitstate \asteps \astato \asteps \astat' \astep 
  \astat$,
  where $\astato = \coren{\astat'}$. 
  But $\astat'$ is not a \daceval{} exit (it is an \daapply{} state), so by the second
  branch of the definition of \dcoren{} we get $\astato = \coren{\astat}$. \\
  By \ih{}, $(\atol{\astato}, \atol{\astat'})$ is entered in \seen{} and
  \work{}; and examined at line 6.
  By lemma~\ref{lem:localsim}, $\atol{\astat} \in \succ{\atol{\astat'}}$ 
  so $(\atol{\astato}, \atol{\astat})$ will be propagated (line 7)
  and entered in \seen{} (line 25).
\item[f)]
  \astat{} is an inner \daceval \\
  This case is simple.
\item[g)]
  \astat{} is a Call \\
  This case is simple.
  \qed
\end{enumerate}
$\phantom{i}$\newline
\begin{thm}[Completeness] ~\\
  After summarization:
  \begin{enumerate}[$\bullet$]
  \item
    For each $(\lstato, \lstatw)$ in \seen,
    there exist \astato, \astatw{} and $p$ such that 
    $p \equiv \ainitstate \asteps \astato \asteps \astatw$ and
    $\lstato = \atol{\astato}$ and
    $\lstatw = \atol{\astatw}$ and
    $\astato \in \corens{\astatw}$
  \item
    For each \lstat{} in \finals, 
    there exist \astat{} and $p$ such that 
    $p \equiv \ainitstate \astepso \astat$ and $\lstat = \atol{\astat}$ and
    \astat{} is a final state.
  \end{enumerate}
\end{thm}
\proof By induction on the number of iterations. 
We prove that the algorithm maintains the following properties for
\seen{} and \finals{}.
\begin{enumerate}
\item \label{thm:sum/tion-complete,prop-non-final}
  For each $(\lstato, \lstatw)$ in \seen, 
  there exist \astato, \astatw{} and $p$ such that 
  $p \equiv \ainitstate \asteps \astato \asteps \astatw$ and
  $\lstato = \atol{\astato}$ and $\lstatw = \atol{\astatw}$ and,
  if \lstatw{} is a \dlceval{} exit then $\astato \in \corens{\astatw}$
  otherwise $\astato = \coren{\astatw}$
\item \label{thm:sum/tion-complete,prop-final}
  For each \lstat{} in \finals, there exist \astat{} and $p$ such that
  $p \equiv \ainitstate \astepso \astat$ and $\lstat = \atol{\astat}$ and
  \astat{} is a final state.
\end{enumerate}
Initially, we must show that the properties hold before the first iteration 
(at the beginning of the algorithm):
\finals{} is empty and \work{} contains just $(\linitstate, \linitstate)$,
for which property 1 holds.

Now the inductive step: at the beginning of each iteration, we remove an edge
$(\lstato, \lstatw)$ from \work{}.
We assume that the properties hold at that point.
We must show that, after we process the edge, the new elements of \seen{} and 
\finals{} satisfy the properties.
\begin{enumerate}[$\bullet$]
\item 
  \lstatw{} is an entry, a \dlcapply{} or an inner \dlceval \\
  $(\lstato, \lstatw)$ is in \seen, so by \ih \\
  $\exists \: \astato, \astatw, p. \;
  p \equiv \ainitstate \asteps \astato \asteps \astatw 
  \;\land\; \lstato = \atol{\astato} 
  \;\land\; \lstatw = \atol{\astatw}
  \;\land\; \astato = \coren{\astatw}$ \\
  For each \lstath{} in \succ{\lstatw}, $(\lstato, \lstath)$ will be propagated.
  \\
  If $(\lstato, \lstath)$ is already in \seen{} then property 
  \ref{thm:sum/tion-complete,prop-non-final} holds by \ih{}
  (in the following cases, we won't repeat this argument and will assume that
  the insertion in \seen{} happens now).
  \\
  Otherwise, we insert the edge at this iteration, at line 25.
  By lemma \ref{lem:localsim-converse}, \\
  $\exists \: \astath. \;
  \lstath = \atol{\astath} \;\land\; \astatw \astep \astath$ \\
  By the second branch of the definition of \dcoren{},
  $\astato = \coren{\astath}$
\item
  \lstatw{} is a call \\
  Let $\lstato = (\denot{\nlam{1}{u_1 \, k_1}{\mcall_1}}, \auarg_1, \henv_1)$ 
  and $\lstatw = (\denot{\slp{}f_2 \: e_2 \: \nlam{2}{u_2}{\mcall_2}\srp^{l_2}}, 
  \tfenv_2, \henv_2)$ \\
  Also, assume \instack{l_2, f_2} (the other cases are simpler). \\
  $(\lstato, \lstatw)$ is in \seen, so by \ih \\
  $\exists \: \astato, \astatw, p. \;
  p \equiv \ainitstate \asteps \astato \astepso \astatw 
  \;\land\; \lstato = \atol{\astato} 
  \;\land\; \lstatw = \atol{\astatw}
  \;\land\; \astato = \coren{\astatw}$ \\
  Each entry \lstath{} in \succ{\lstatw} will be propagated.
  By lemma \ref{lem:localsim-converse}, \\
  $\exists \: \astath. \;
  \lstath = \atol{\astath} \;\land\; \astatw \astep \astath$ \\
  Since $\astath = \coren{\astath}$, 
  property \ref{thm:sum/tion-complete,prop-non-final}
  holds for \lstath. \\
  If there is no edge $(\lstath, \lstatf)$ in \summary{}, we are done. \\
  Otherwise, we call \tw{Update(}\lstato, \lstatw, \lstath, \lstatf\srp{}
  and we must show that property \ref{thm:sum/tion-complete,prop-non-final}
  holds for the edge inserted in \seen{} by \tw{Update}.
  \\
  Let $\stenv_1$ be the stack of \astato{}.
  By lemma \ref{lem:same-level}, the stack of \astatw{} is 
  $\tfenv_2 :: \stenv_1$. 
  \\
  Let $\lstath = (\denot{\nlam{3}{u_3 \, k_3}{\mcall_3}}, \auarg_3, \henv_2)$
  and $\lstatf = (\denot{\slp{}k_4 \: e_4\srp^{l_4}}, \tfenv_4, \henv_4)$. \\
  (Note that $\tfenv_4$ contains only user bindings.) \\
  We know $\summary \subseteq \seen$ so by \ih{} for $(\lstath, \lstatf)$ 
  we get (note that \lstatf{} is a \dlceval{} exit) \\
  $\exists \: \astath', \astatf', p'. \;
  p' \equiv \ainitstate \asteps \astath' \astepso \astatf' 
  \;\land\; \lstath = \atol{\astath'} 
  \;\land\; \lstatf = \atol{\astatf'}
  \;\land\; \astath' \in \mi{CE}^*_{p'}(\astatf')$ \\
  Then, $\astath' = (\denot{\nlam{3}{u_3 \, k_3}{\mcall_3}}, \auarg_3, 
  \acarg_3, \stenv_3', \henv_2)$ and by lemma \ref{lem:same-level}, \\
  $\astatf' = (\denot{\slp{}k_4 \: e_4\srp}, 
  \tfenv_4\onemap{k_4}{\acarg_3}::\stenv_3', \henv_4)$. \\
  But the path from $\astath'$ to $\astatf'$ is push monotonic, so by lemma
  \ref{lem:stack-irrel} there exist states \\
  $\astath = (\denot{\nlam{3}{u_3 \, k_3}{\mcall_3}}, \auarg_3, 
  \denot{\nlam{2}{u_2}{\mcall_2}}, \stenv_3, \henv_2)$ \\
  where $\stenv_3 = 
  \tfenv_2\onemap{f_2}{\mset{\denot{\nlam{3}{u_3 \,k_3}{\mcall_3}}}}::\stenv_1$,
  and $\astatf = (\denot{\slp{}k_4 \: e_4\srp}, \stenv_4, \henv_4)$ \\
  where $\stenv_4 = 
  \tfenv_4\onemap{k_4}{\denot{\nlam{2}{u_2}{\mcall_2}}}::\stenv_3$, 
  such that $\astath{} \astepso \astatf{}$. \\
  Thus, the path $p$ can be extended to $\ainitstate \asteps \astato \astepso 
  \astatw \astep \astath \astepso \astatf$.
  By the abstract semantics, the successor \astat{} of \astatf{} is 
  $(\denot{\nlam{2}{u_2}{\mcall_2}}, \ubiga{e_4, l_4, \stenv_4, \henv_4}, 
  \stenv_3, \henv_4)$. \\
  The state \lstat{} produced by \tw{Update} is 
  $(\denot{\nlam{2}{u_2}{\mcall_2}}, \aubiga{e_4, l_4, \tfenv_4, \henv_4}, 
  \tfenv, \henv_4)$ where $\tfenv = 
  \tfenv_2\onemap{f_2}{\mset{\denot{\nlam{3}{u_3 \, k_3}{\mcall_3}}}}$.
  It is simple to see that $\lstat = \atol{\astat}$.
\item
  \lstatw{} is a \dlceval{} exit, 
  $(\denot{\slp{}k \: e\srp^{l_2}}, \tfenv_2, \henv_2)$ \\
  If \lstato{} is \linitstate{} then \tw{Final(}\lstatw\srp{} is called
  and a local state \lstat{} of the form \\
  $(\haltcont, \aubiga{e, l_2, \tfenv_2, \henv_2}, \emptyset, \henv_2)$
  goes in \finals.
  We must show that property \ref{thm:sum/tion-complete,prop-final} holds. \\
  By \ih{} for $(\lstato, \lstatw)$,
  $\quad \exists \: \astatw, p. \;
  p \equiv \ainitstate \astepso \astatw 
  \;\land\; \lstatw = \atol{\astatw}
  \;\land\; \ainitstate \in \corens{\astatw}$. \\
  (Note that $\astato = \ainitstate$.)
  By lemma \ref{lem:same-level}, the stack $\stenv_2$ of \astatw{} is
  $\tfenv_2\onemap{k}{\haltcont}::\tuple{}$.
  Hence, the successor \astat{} of \astatw{} is 
  $(\haltcont, \ubiga{e, \stenv_2, \henv_2}, \tuple{}, \henv_2)$,
  and $\lstat = \atol{\astat}$ holds.
  \\
  If $\lstato \neq \linitstate$,
  for each triple $(\lstath, \lstatf, \lstato)$ in \callers{}, 
  we call \tw{Update(}$\lstath, \lstatf, \lstato, \lstatw$\srp{}.
  Insertion in \callers{} happens only at line 11, which means that
  $(\lstath, \lstatf)$ is in \seen.
  Thus, by \ih{} \\
  $\exists \: \astath, \astatf, p. \;
  p \equiv \ainitstate \asteps \astath \astepso \astatf
  \;\land\; \lstath = \atol{\astath}
  \;\land\; \lstatf = \atol{\astatf}
  \;\land\; \astath = \coren{\astatf}$ \\
  Also, $\lstatf \lstep \lstato$ thus by lemma \ref{lem:localsim-converse}
  $\exists \: \astato. \; 
  \astatf \astep \astato
  \;\land\; \lstato = \atol{\astato}$ \\
  Using the \ih{} for $(\lstato, \lstatw)$ and lemma \ref{lem:stack-irrel}
  we can show that the edge inserted by \tw{Update} satisfies 
  property \ref{thm:sum/tion-complete,prop-non-final} (similar to the previous
  case).
  \\
  For each triple $(\lstath, \lstatf, \lstato)$ in \tcallers{}, 
  we call \tw{Propagate(}$\lstath, \lstatw$\srp{}.
  We must show that property \ref{thm:sum/tion-complete,prop-non-final} holds
  for $(\lstath, \lstatw)$.
  Insertion in \tcallers{} happens only at line 23, which means that
  $(\lstath, \lstatf)$ is in \seen.
  By \ih{} for $(\lstato, \lstatw)$ and $(\lstath, \lstatf)$ and by 
  lemma \ref{lem:stack-irrel},
  we can show that there are states \astath{} and \astatw{} and path $p'$
  such that $\lstath = \atol{\astath}$, $\lstatw = \atol{\astatw}$
  and $\astath \in \mi{CE}^*_{p'}(\astatw)$.
  Hence, property \ref{thm:sum/tion-complete,prop-non-final} holds for
  $(\lstath, \lstatw)$.
\item
  \lstatw{} is a tail call \\
  Similarly.
  \qed
\end{enumerate}

\end{document}